\documentclass[twocolumn,showpacs,preprintnumbers,amsmath,amssymb]{revtex4}

\usepackage[latin1, ansinew]{inputenc}
\usepackage{graphicx}
\usepackage{dcolumn}
\usepackage{bm}
\usepackage[T1]{fontenc}
\usepackage[dvips]{epsfig}
\usepackage{epsf}
\usepackage{color}

\begin{document}

\title{Annular shear of cohesionless granular materials:\\
from inertial to quasistatic regime}

\author{Georg Koval, Jean-Noël Roux, Alain Corfdir and François Chevoir}
\email{chevoir@lcpc.fr}
\affiliation{Université Paris-Est,
Institut Navier, Champs sur Marne, France}

\date{\today}

\begin{abstract}

Using discrete simulations, we investigate the behavior of a model
granular material within an annular shear cell. Specifically,
two-dimensional assemblies of disks are placed between two
circular walls, the inner one rotating with prescribed angular
velocity, while the outer one may expand or shrink and maintains a
constant radial pressure. Focusing on steady state flows, we
delineate in parameter space the range of applicability of the
recently introduced constitutive laws for sheared granular
materials (based on the inertial number). We discuss the two
origins of the stronger strain rates observed near the inner
boundary, the vicinity of the wall and the heteregeneous stress
field in a Couette cell.  Above a certain velocity, an inertial
region develops near the inner wall, to which the known
constitutive laws apply, with suitable corrections due to wall
slip, for small enough stress gradients. Away from the inner wall,
slow, apparently unbounded creep takes place in the nominally
solid material, although its density and shear to normal stress
ratio are on the jammed side of the critical values. In addition
to rheological characterizations, our simulations provide
microscopic information on the contact network and velocity
fluctuations that is potentially useful to assess theoretical
approaches.

\end{abstract}


\maketitle

\section{Introduction} \label{sec:intro}

Significant progress in the modeling of dense granular flow in the
inertial regime has been brought about by the recently introduced
viscoplastic laws~\cite{Pouliquen02a,Gdr04,Forterre08}, as
identified in experiments and discrete numerical simulations in
two-dimensional (2D)~\cite{Dacruz05, Lois05,Chevoir08b} and
three-dimensional (3D)~\cite{Jop06,Hatano07,Peyneau08} situations.

One typically considers homogeneous assemblies of grains of size
$d$ and mass density $\rho_p$, under shear stress $\sigma$ and
average pressure $P$. Denoting the shear rate as $\dot\gamma$,
constitutive laws are conveniently expressed as relations between
dimensionless quantities: effective friction  $\mu^*$
($=\sigma/P$), solid fraction $\nu$, and most noticeably  inertial
number $I= \dot \gamma d \sqrt{\rho_p/P}$, thus rescaling  various
experimental data into a consistent picture. As the ratio of the
inertial to shear times, the latter parameter quantifies the
inertial effects. For a frictional material, a small value of $I$
($\le 10^{-2}$) corresponds to the quasistatic critical state
regime, while a large value of $I$ ($\ge 10^{-1}$) corresponds to
the collisional regime~\cite{Goldhirsch03}. As $I$ increases,
solid fraction $\nu$ decreases approximately linearly starting
from a maximum value $\nu_{max}=\nu_c$ (\emph{dynamic dilatancy
law}), while the effective friction coefficient $\mu^*$ increases
approximately linearly starting from a minimum value
$\mu^*_{min}=\tan \phi$ (\emph{dynamic friction law}). This yields
a viscoplastic constitutive law, with a Coulomb frictional term
and a Bagnold viscous term.

In the quasistatic regime ($I \to 0$) this approach indicates
$\mu^* \to \mu^*_{min}$ constant, independently on the strain, in
steady shear flow. Below this minimum stress ratio, quasistatic
strains are possible that are described by elastoplastic models.
For large enough shear strains~\cite{Schofield68,Wood90}, a
solidlike material approaches the so-called critical state, which
coincides with the state of steady shear flow in the limit of
$I\to 0$.

Once constitutive laws are obtained on dealing with homogeneous
systems, they should be locally applicable to all possible flow
geometries. Of course, they are quite unlikely to provide a proper
description of some strongly heterogeneous situations occurring
when strain is localized near boundaries, in thin layers, on a
scale of a few grains. Yet, for smoothly varying stress fields,
they might prove sucessful, as was shown e.g., with flows down
inclined planes. Those were studied in the absence of lateral
walls both experimentally and through discrete simulations
(see~\cite{Gdr04} for a review). Then the stress distribution
becomes heterogeneous but the effective friction remains constant,
so that the situation is comparable to homogeneous flows. More
remarkably, a three dimensional version of the constitutive
law~\cite{Jop06,Deryck08a} was found to model similar flows
between lateral walls, which induce truly three-dimensional stress
distributions and velocity profiles~\cite{Jop05b}.

Other simple geometries are the vertical chute and the annular
shear~\cite{Gdr04}. The present paper investigates the material
behavior in the annular (\emph{Couette}) shear geometry, for which
the sample is confined between two rough cylinders and sheared by
the rotation of the inner one. The annular shear cell is a
classical experimental device to measure the rheological
properties of complex fluids, and has been used for granular
materials, both in two dimensions~\cite{Miller96, Elliott98,
Veje99, Howell99a, Hartley03} and in three
dimensions~\cite{Tardos98, Howell99b, Mueth00, Klausner00, Cain01,
Losert01, Bocquet02b, Dacruz02, Mueth03, Chambon03a, Tardos03,
Utter04a, Dacruz04a, Daniel07, Wang08}. In this geometry, the
stress distribution is well known, as will be detailed in the
following: the normal stress is approximately constant while the
shear stress strongly  decreases away from the inner wall. The
decrease of $\mu^*$ away  from the inner wall then explains the
localization of the shear. We may even expect a transition between
an inertial flow near the inner wall, where $\mu^* \ge
\mu^*_{min}$, and a quasistatic regime further, which analysis
would help understanding shear localization near a wall (influence
of shear rate and confining pressure on the width and dilation of
shear bands), of interest in industrial
conducts~\cite{Nedderman92}, geotechnical
situations~\cite{Savage89b} and tectonophysics~\cite{Chambon06b}.

Following previous discrete simulations~\cite{Karion99,
Schollmann99, Latzel00, Zervos00}, we investigate the rheology and
the microstructure of granular materials in this geometry. We
consider two-dimensional, slightly polydisperse assemblies of
cohesionless frictional disks. This allows to vary the shear state
and provides access to microscopic information at the scale of the
grains and of the contact network, hardly measurable
experimentally. We prescribe the shear rate and the pressure,
allowing global dilation of the shear cell. To save computation
time, we implement periodic boundary conditions. All along this
paper, we shall compare our results with the homogeneous shear
case~\cite{Dacruz05}.

Sec.~\ref{sec:system} is devoted to the description of the
simulated system, its preparation and the definition of
dimensionless control parameters. Sec.~\ref{sec:localization}
describes the influence of the shear velocity and of the system
scale on the shear localization near the inner wall through the
radial profiles of various quantities. Sec.~\ref{sec:behavior}
shows the validity of the previous constitutive law for inertial
regime, and analyzes its limit in quasistatic regime.
Sec.~\ref{sec:prediction} then explains how the constitutive law
is able to predict various quantities measured in
Sec.~\ref{sec:localization}.

Preliminary and complementary results are presented
in~\cite{Koval08a}.

\section{Simulated systems} \label{sec:system}

\subsection{Annular shear}

\begin{figure}[!htb]
\begin{center}
\includegraphics*[width=8cm]{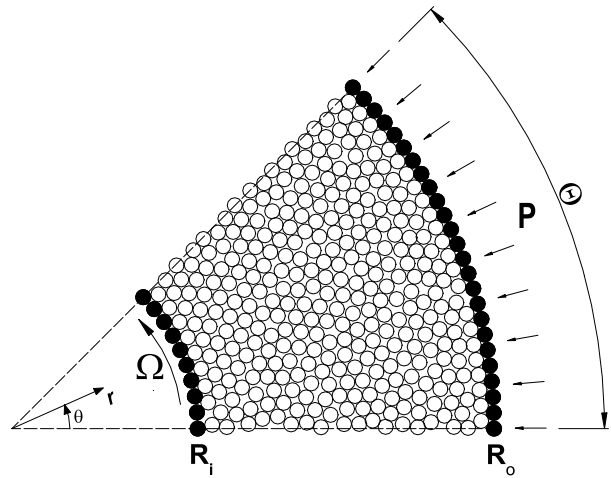}
\caption{\label{Fig1} \textit{Annular shear geometry (black grains
constitute the rough walls).}}
\end{center}
\end{figure}

The simulated systems are two dimensional (2D) as indicated in
Fig.~\ref{Fig1}. The granular material is a dense assembly of $n$
dissipative disks of average diameter $d$ and average mass $m$. A
small polydispersity of $\pm 20 \%$ prevents crystallization.

The granular material is subjected to annular shear between two
circular rough walls. The outer wall (radius $R_o$) does not
rotate, while the inner wall (radius $R_i$) moves at the
prescribed rotation rate $\Omega$. The wall roughness, which
reduces sliding, is made of contiguous \emph{glued} grains with
the same characteristics as the flowing grains (polydispersity and
mechanical properties). We call $r$ and $\theta$ the radial and
orthoradial directions. $r = R_i$ and $r = R_o$ respectively
correspond to the centers of the grains which compose the inner
and outer walls . As the grains of the inner wall form one rigid
body, the motion of each of them combines a translation of its
center with velocity $\Omega R_i \vec e_{\theta}$ and a rotation
rate $\Omega$.

Since we have not observed any influence of the outer wall on the
behavior of the sheared material close to the inner wall for
$R_{o} \geq 2 R_{i}$, we have set $R_{o}= 2 R_{i}$, in the results
presented in this paper. The geometry is then defined by the sole
value of $R_{i}/d$.

An important feature of our discrete simulations is the control of
the normal stress exerted by the outer wall on the grains, as done
in~\cite{Zervos00}. We prescribe $\sigma_{rr}(R_{o})=P$, through
the radial motion of the outer wall, given by: $\dot R_o = (P -
\sigma_{rr}(R_{o}))/ g_{p}$, where $g_p$ is a viscous damping
parameter. In steady state, the motion of the outer wall
oscillates around a mean value ($\langle \dot R_o \rangle =0$)
corresponding to a prescribed value of the normal stress at this
point ($\langle \sigma_{rr}(R_{o}) \rangle = P$). Such control of
the radial stress is applied in \emph{cylinder shear apparatus}
aimed at studying the behavior of soils near an
interface~\cite{Corfdir04}. It differs from most experiments and
discrete simulations, where the volume is fixed in two
dimensions~\cite{Elliott98, Howell99a, Karion99, Latzel00,
Latzel03, Utter04a}, or dilatancy is possible through the free
surface in three dimensions ~\cite{Tardos98, Mueth00, Cain01,
Losert01, Bocquet02b, Mueth03, Tardos03, Dacruz04a, Daniel07,
Wang08}.

We use the standard \emph{spring-dashpot} contact law described
in~\cite{Dacruz05}, which introduces the coefficients of
restitution $e$ and friction $\mu$, and the elastic stiffness
parameters $k_n$ and $k_t$. Discrete simulations are carried out
with standard molecular dynamics method, as
in~\cite{Cundall79,Silbert01,Roux05,Dacruz05,Rognon08a,Peyneau08}.
The equations of motion are discretized using Gear's order three
predictor-corrector algorithm~\cite{Allen87}.

To decrease the computation time, we have introduced periodic
boundary condition along $\theta$, exploiting the angular
invariance~\cite{Cui07}. This reduces the representation of the
annular shear cell to an angular sector $0 \leq \theta \leq
\Theta$ ($\theta < \pi$) instead of the whole system $0 \leq
\theta \leq 2\pi$. $\Theta = 2 \pi/N$, where $N$ is an integer.
The description of this method, together with the analysis used to
choose the values of $\Theta$ according to the size of the
systems, is presented in App.~A. The list of simulated geometries
is given in Tab.~I. We notice that the studied systems are much
larger than in previous discrete simulations.

\begin{table}[!htb]
    \begin{center}
    \caption{\textit{List of simulated geometries.}}
    \begin{tabular}{c c c c c c}
        \\
        \hline
          &  & $n$ & $R_i/d$ & $\Theta /2 \pi$ \\
        \hline
        & $R_{25}$ & $1500$ & $25$ & $1/4$ \\
        \hline
        & $R_{50}$ & $3100$ & $50$ & $1/8$ \\
        \hline
        & $R_{100}$ & $8000$ & $100$ & $1/12$ \\
        \hline
        & $R_{200}$ & $15700$ & $200$ & $1/24$ \\
    \end{tabular}
    \end{center}
\end{table}

\subsection{Dimensional analysis}

In our discrete simulations, the system is completely described by
a list of independent parameters associated to the grains and to
the shear state. As a way to reduce the number of parameters, it
is convenient to use dimensional analysis, which guarantees that
all the results can be expressed as relations between
dimensionless quantities.

The grains are described by their size $d$ and mass $m$, their
coefficients of restitution $e$ and friction $\mu$, and their
elastic stiffness parameters $k_n$ and $k_t$. It was shown in
previous discrete simulations~\cite{Silbert01, Campbell02,
Dacruz05}, that $k_t/k_n$ and $e$  have nearly no influence on
dense granular flows. Consequently, $k_t/k_n$ was fixed to $0.5$
and $e$ was fixed to $0.1$. The influence of $\mu$, especially
near $\mu = 0$, has been shown in~\cite{Dacruz05}. In this paper,
we restrict our analysis to the value $\mu = 0.4$, except for the
discussion of the constitutive law where the case of frictionless
grains ($\mu = 0$) will be also analyzed. Results for other values
of $\mu$ may be found in~\cite{Koval08a}.

The shear state is described by the prescribed normal stress on
the outer wall $P$, the rotation rate of the inner wall $\Omega$,
the radius $R_i$ and $R_o$ of the two walls, and the viscous
damping parameter $g_p$. We have not observed any influence of
$g_p$, once the shear zone is localized near the inner wall and
separated by a relatively thick layer of material from the outer
wall. The dimensionless number $g_{p}/\sqrt{mk_{n}}$ remains of
order $0.1$ in all our simulations, so that the time scale of the
fluctuations of $R_o$ is imposed by the material rather than the
wall, and that the wall \emph{sticks} to the material.
Consequently, the shear state is described by the geometric
parameters $R_i/d$ ($R_o=2R_i$), and by the dimensionless
tangential velocity of the inner wall (also called shear
velocity):

\begin{eqnarray}
\label{eqn:Vtheta} V_{\theta} = \frac{\Omega R_i}{d}
\sqrt{\frac{m}{P}},
\end{eqnarray}

\noindent which is similar to the notion of \emph{inertial
number}, but at the scale of the whole system. A small value of
$V_{\theta}$ corresponds to the quasistatic regime, while a large
value corresponds to the collisional regime. Seven values of
$V_{\theta}$ have been studied systematically for all systems:
$0.0025$, $0.025$, $0.25$, $0.5$, $1.0$, $1.5$ and $2.5$. The
value $0.00025$ was also considered in a few cases.

Moreover the stress scales $k_n$ and $P$ may be compared through
the dimensionless number $\kappa = k_n/P$. Let us call $h$ the
normal deflection of the contact (or apparent
\emph{interpenetration} of undeformed disks). Being inversely
proportional to the relative deflection $h/d$ of the contacts for
a confining stress $P$, $\kappa$ is called \emph{contact stiffness
number}~\cite{Dacruz05}. A large value corresponds to rigid
grains, while a small value corresponds to soft grains. It was
shown~\cite{Dacruz05} that it has no influence on the results once
it exceeds $10^4$, which is the value chosen in all our discrete
simulations (\emph{rigid grain limit}).

In the following (both text and figures), the length, mass, time
and stress are made dimensionless by $d$, $m$, $\sqrt{m/P}$ and
$P$, respectively.

Table~II gives the list of material parameters.

\begin{table}[!htb]
    \caption{\textit{List of material parameters.}}
    \begin{tabular}{c c c c c c}
        \\
        \hline
          & polydispersity & $\mu $ & $e$ & $k_t/k_n$ & $\kappa$ \\
        \hline
        & $\pm 20 \%$ & $0.4$ & $0.1$ & $0.5$ & $10^4$\\
        \hline
    \end{tabular}
\end{table}

\subsection{Steady shear states}

For a given sample, the first step consists in depositing the
grains without contact and at rest between the two distant walls.
Applying a normal stress at the outer wall, we compress the
assembly of grains, considering first that they are frictionless
($\mu=0$), so as to get a very dense initial state. Except near
the walls, its solid fraction is close to $0.85$, near the random
close packing of slightly polydispersed disks~\cite{Combe02a}.
When the granular material supports completely the applied normal
stress, the grains are at rest and the dense system is ready to be
sheared. We start to shear the material (now considering that the
grains are frictional) imposing the rotation of the inner disk.
After a transient, the system reaches a steady state,
characterized by constant time-averaged profiles of solid
fraction, velocity and stress. In practice, the stabilization of
the profiles depends on the considered variable. If we take the
inner wall displacement $V_{\theta} \Delta t$ (where $\Delta t$ is
the simulation time) as a shear length parameter, the stresses
usually present a short transient on a distance around $V_{\theta}
\Delta t \leq 5$. However, the stabilization of the solid fraction
rather requires $V_{\theta} \Delta t \approx 50$, mostly because
of the very dense initial state. Consequently we consider that the
condition to reach a steady state is $V_{\theta} \Delta t \geq 100
$. This procedure provides an initial state with a shear velocity
$V_{\theta}$. As a way to guarantee an initial state consistent
for the comparisons between discrete simulations with different
$V_{\theta}$, the procedure is first applied with the highest
value of $V_{\theta}$, and then $V_{\theta}$ is progressively
decreased.

In steady state, we consider that the statistical distribution of
the quantities of interest (structure, velocities, forces\dots)
are independent of $t$ and $\theta$, so that we average both in
space (along $\theta$) and in time (considering $200$ time steps
distributed over the distance $V_{\theta} \Delta t \geq 200$).
Then we calculate the profiles of solid fraction, velocity and
stress components according to the averaging procedure described
in App.~B.

Beyond the number of acquisition points, the consistency of the
averaged values depends on the shear strain accumulated during the
acquisition of data. We concentrate our interest on the region
where the system may be considered in a steady state, which occurs
at large enough shear strain. Based on the observation of the
transients, we consider that this is true when $\dot \gamma \Delta
t \geq 10$. Because of the strain localization, the region of
interest is located near the inner wall and limited to $R_i \to
R_i + R_{steady}$ where the value of $R_{steady}$ is given in
Tab.~III.

\begin{table}[!htb]
    \begin{center}
    \caption{\textit{Limit of the steady state region.
    Minimum and maximum values correspond to global quasistatic regime and to $V_{\theta}=2.5$ respectively.}}
    \begin{tabular}{c c c}
        \\
        \hline
          &  & $R_{steady}$  \\
        \hline
        & $R_{25}$ & $7-17$  \\
        \hline
        & $R_{50}$ & $9-25$  \\
        \hline
        & $R_{100}$ & $13-35$ \\
        \hline
        & $R_{200}$ & $18-52$ \\
    \end{tabular}

    \end{center}
\end{table}

\section{Localized shear states} \label{sec:localization}

In this section, we show the shear localization near the inner
wall through the radial profiles of different quantities. In
App.~C we focus on internal variables associated to the contact
network~\cite{Radjai04,Radjai08} (coordination number $Z$ and
mobilization of friction $M$) and to the fluctuations of the
motion of the grains, translational or rotational. We
systematically discuss the influence of $V_{\theta}$.

\subsection{Stress field} \label{sec:stress}

In steady ($\frac{\partial}{\partial t} = 0$) annular shear flows
($\frac{\partial}{\partial \theta} = 0$), without radial flow
($v_r = 0$), continuum mechanics predicts~\cite{Coleman66} a
variation of normal stress $\sigma_{rr}$ related to the velocity
profile, and a $1/r^2$ decrease of the shear stress $\sigma_{r
\theta}$ associated to the conservation of the torque:

\begin{eqnarray}
\label{eqn:stressa}
  \frac{4\nu}{\pi}  \frac{v_{\theta}^2}{r} &=& \frac{\partial \sigma_{rr}}{\partial
  r} +  \frac{\sigma_{rr} - \sigma_{\theta \theta}}{r},
\end{eqnarray}

\begin{eqnarray}
\label{eqn:stressb}
  \sigma_{r \theta} &=& S (\frac{R_i}{r})^2,
\end{eqnarray}

\noindent where $\nu(r)$ and $v_{\theta}(r)$ are the solid
fraction and orthoradial velocity profiles, $S$ is the shear
stress at the inner wall ($S=\sigma_{r \theta}(R_i)$) and
$\sigma_{ii}$ are positive for compression.

\begin{figure}[!htb]
\begin{center}
\includegraphics*[width=7cm]{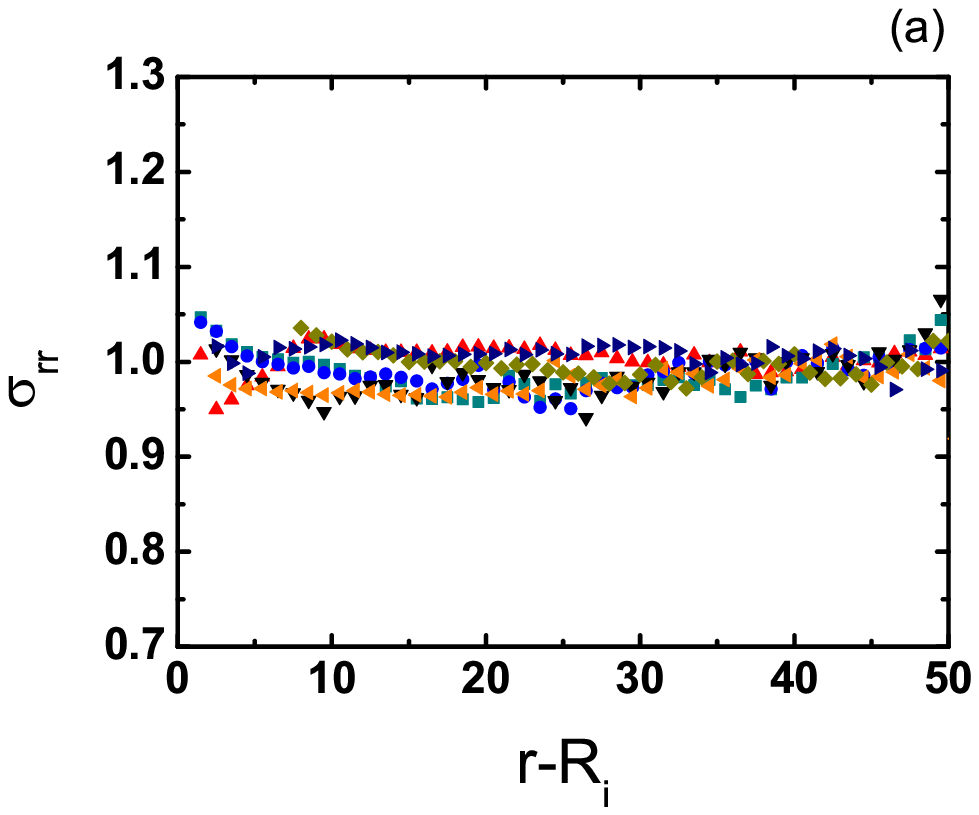}
\includegraphics*[width=7cm]{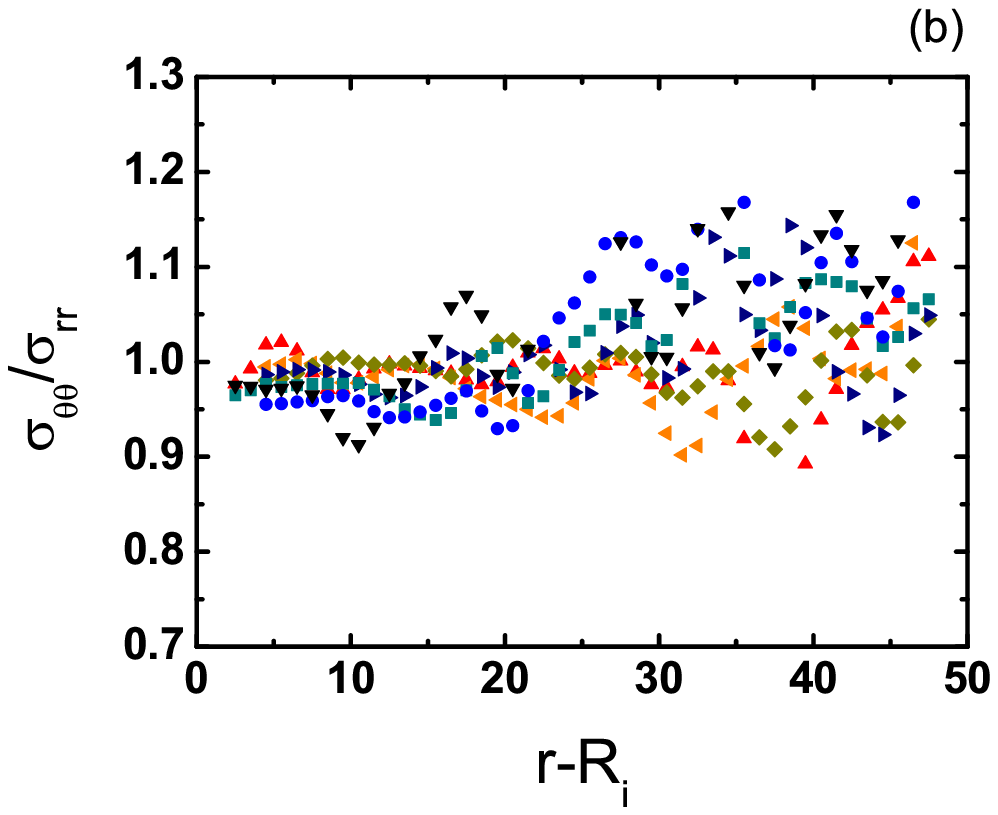}
\caption{\label{Fig2} \textit{(Color online) (a) Normal stress
$\sigma_{rr}(r)$ and (b) ratio between the normal and the
orthonormal stresses $\sigma_{\theta \theta}/\sigma_{rr}(r)$
profiles for different shear velocities. ($\blacktriangledown$)
$V_{\theta}=0.0025$, ($\textcolor[rgb]{0.00,0.00,1.00}{\bullet}$)
$V_{\theta}=0.025$,
($\textcolor[rgb]{0.25,0.50,0.50}{\blacksquare}$)
$V_{\theta}=0.25$,
($\textcolor[rgb]{0.00,0.00,0.50}{\blacktriangleright}$)
$V_{\theta}=0.5$,
($\textcolor[rgb]{0.50,0.50,0.00}{\blacklozenge}$)
$V_{\theta}=1.0$,
($\textcolor[rgb]{1.00,0.50,0.00}{\blacktriangleleft}$)
$V_{\theta}=1.5$,
($\textcolor[rgb]{0.98,0.00,0.00}{\blacktriangle}$)
$V_{\theta}=2.5$. Geometry $R_{50}$.}}
\end{center}
\end{figure}

Fig.~\ref{Fig2}a shows the coarse-grained profiles of the normal
stress component $\sigma_{rr}$ in geometry $R_{50}$ for different
wall velocities $V_{\theta}$, while Fig.~\ref{Fig2}b shows the
ratio between the orthonormal and the normal stresses
$\frac{\sigma_{rr}}{\sigma_{\theta \theta}}$. The normal stress
$\sigma_{rr}$ is nearly constant and equal to the confining
pressure $P$. The $\partial \sigma_{rr}/\partial r$ term in the
momentum equation~\eqref{eqn:stressa} smoothes the $\sigma_{rr}$
profile, which might explain the absence of fluctuations of
$\sigma_{rr}$. A crude estimate of the centrifugal effects may be
given, if the last term of equation~\eqref{eqn:stressa} is
neglected, and, anticipating on Sec.~\ref{sec:velocity} and
Sec.~\ref{sec:compa}, a constant solid fraction $\nu \simeq 0.8$
is assumed and an exponential velocity profile
$v_{\theta}(r)=V_{\theta} \exp\left(-(r-R_i)/\ell\right)$, with
$\ell$ between $2$ and $6$:

\begin{eqnarray}
\mid \sigma_{rr}(R_i)-1\mid \le \frac{2 \pi \nu R_i}{\ell}
V_{\theta}^2.
\end{eqnarray}

\noindent Consequently, for $R_i = 50$, $\mid
\sigma_{rr}(R_i)-1\mid \le 0.05$ for $V_{\theta}=1$ and $\ell =
5$. For $V_{\theta}=2.5$, the centrifugal effects might become
significant, however it has not been observed.

The radial $\sigma_{rr}$ and orthoradial $\sigma_{\theta \theta}$
stresses are nearly equal for $r-R_i \lesssim 10$. This has
already been observed in other configurations (plane
shear~\cite{Dacruz05,Lois05,Peyneau08} within less than $5\%$,
inclined plane~\cite{Silbert01,Dacruz04a,Lois05}), and was
previously reported in annular shear~\cite{Latzel00,Latzel03}.
This very small normal stress difference is not explained yet. The
fluctuations of $\sigma_{\theta \theta}$ for $r-R_i \gtrsim 15$
probably reflect the frozen disorder beyond the steady zone, where
the material is much less deformed than closer to the inner wall,
so that the time averaging is unsufficient. Consequently these
fluctuations increase as $V_{\theta}$ decreases.

\begin{figure}[!htb]
\begin{center}
\includegraphics*[width=7cm]{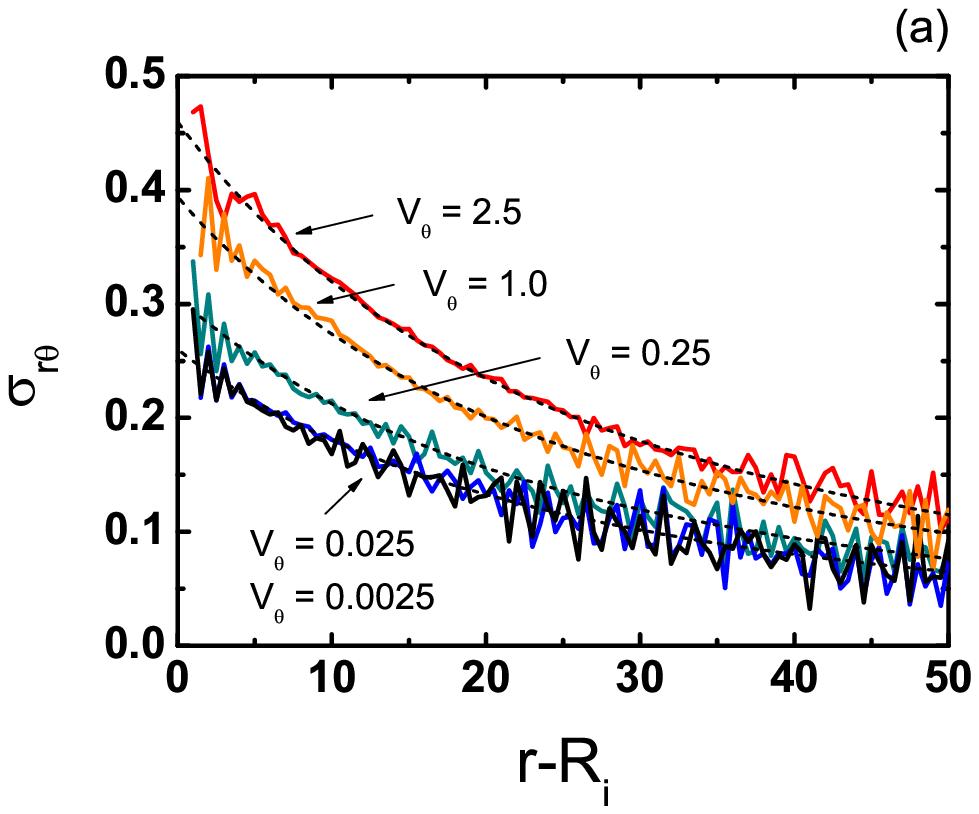}
\includegraphics*[width=7.3cm]{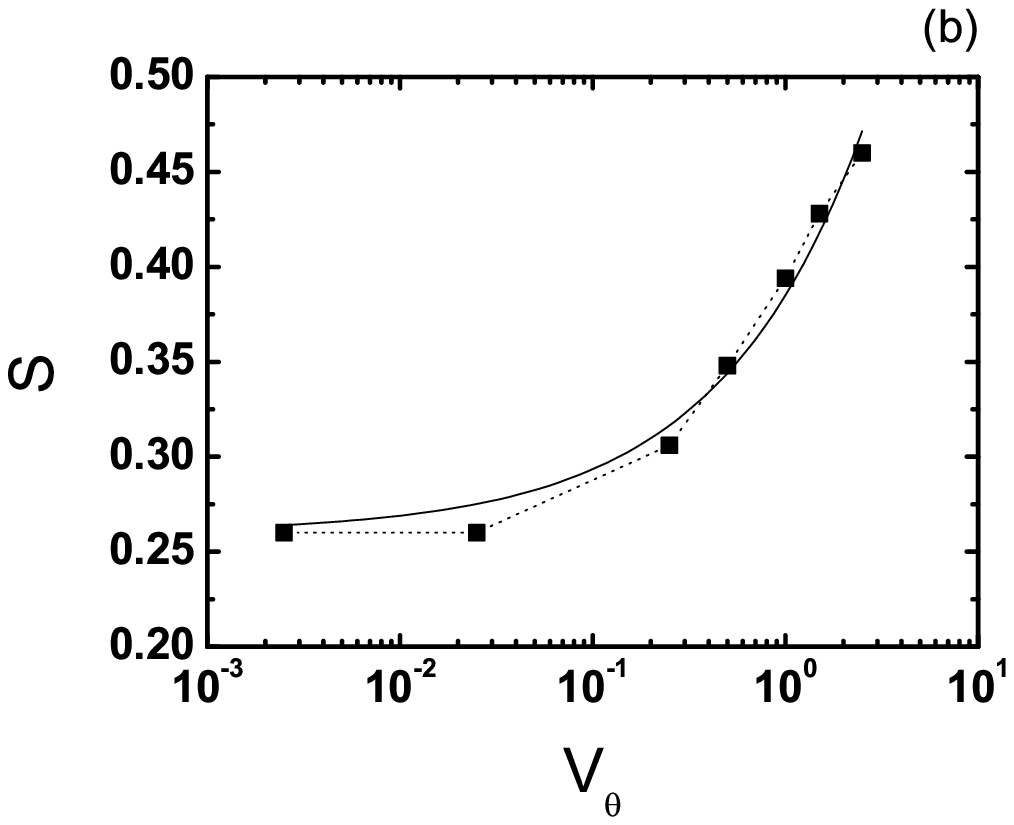}
\caption{\label{Fig3} \textit{(Color online) (a) Shear stress
profiles $\sigma_{r \theta}(r)$ (solid lines) and fit according
to Eq.~\eqref{eqn:stressb} (dashed lines) for different wall
velocities $V_{\theta}$. (b) Shear stress at the inner wall $S$
as function of $V_{\theta}$ (semi-logarithmic scale). The solid
line represents the function: $S=0.26+0.13V_{\theta}^{0.57}$.
Geometry $R_{50}$.}}
\end{center}
\end{figure}

The shear stress profiles $\sigma_{r\theta}(r)$ shown in
Fig.~\ref{Fig3}a (for different shear velocity $V_{\theta}$) are
consistent with the $1/r^2$ decrease of \eqref{eqn:stressb}. The
oscillations about the mean value are due to the material
structuration near the inner wall (see Sec.~\ref{sec:compa}) and
to the frozen disorder in the very slowly sheared regions, which
is beyond the steady zone.

Fig.~\ref{Fig3}b shows the dependence of the shear stress at the
inner wall $S$ on shear velocity $V_{\theta}$. Below a certain
value ($V_{\theta}\lesssim 0.025$), $S$ tends to a finite limit.
Consequently, the shear stress profiles $\sigma_{r\theta}(r)$
become independent of $V_{\theta}$. This behavior characterizes
the global (that is to say, in the whole system) quasistatic
regime, where the stresses (and other state variables) do not
depend on the velocity. However, for $V_{\theta}\gtrsim 0.025$,
inertial effects become significant and $S$ increases with
$V_{\theta}$. Previous works reported a similar dependence of the
shear stress on the shear velocity in other configurations
(see~\cite{Gdr04} for a review). More specifically, the
experimental measurement of the torque as a function of the
rotation rate in the annular shear geometry indicates a transition
from a rate independent to a rate dependent regime~\cite{Savage84,
Tardos98, Klausner00, Dacruz02, Daniel07}. Our results can be
approximated by a function like $S=S_{qs}+\alpha
V_{\theta}^{\beta}$, where $S_{qs}$ is the global quasistatic
limit value, $\alpha$ and $\beta$ are two constants. We notice
that $\beta$ is close to $1/2$ rather than $2$ as might be naïvely
expected from Bagnold's rheology. We notice that in the experiment
of~\cite{Tardos98}, the transition occurs for $V_{\theta} \simeq
0.3$ (after appropriate rescaling), which is not far from what is
observed in Fig.~\ref{Fig3}b. However, we also point out that the
$S(V_{\theta})$ curve, here shown for geometry $R_{50}$, in fact
depends on the geometry.

\subsection{Velocity field} \label{sec:velocity}

\begin{figure}[!htb]
\begin{center}
\includegraphics*[width=7cm]{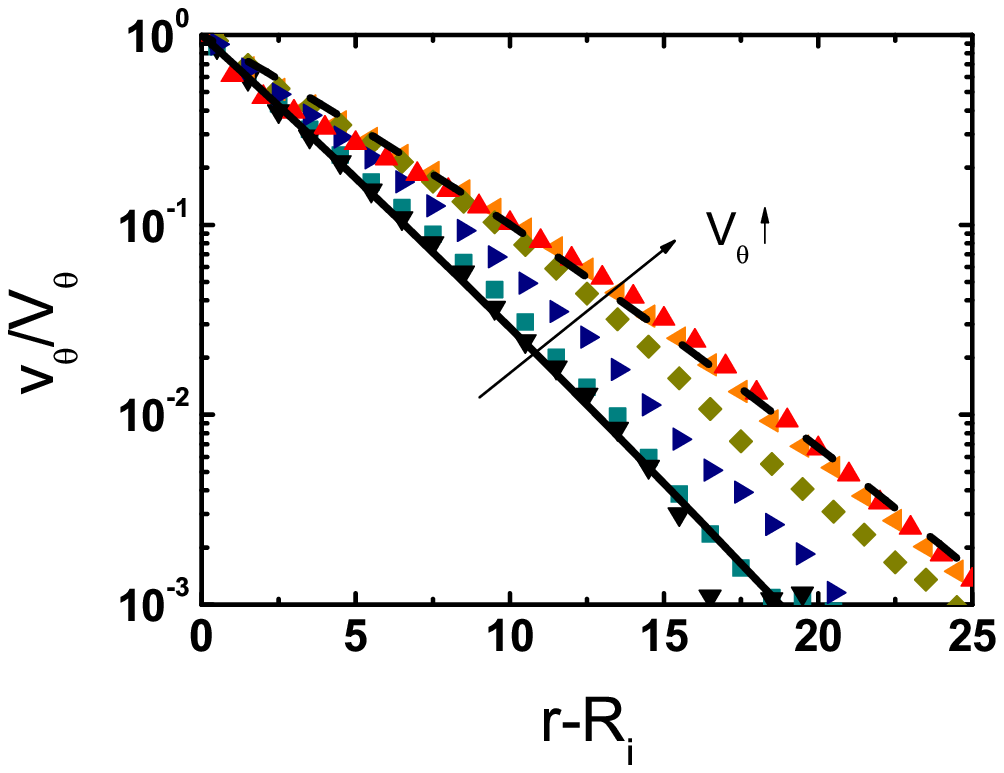}
\caption{\label{Fig4} \textit{(Color online) Influence of the
shear velocity $V_{\theta}$ on the velocity profiles
$v_{\theta}(r)$ (semi-logarithmic scale). ($\blacktriangledown$)
$V_{\theta}=0.0025$,
($\textcolor[rgb]{0.25,0.50,0.50}{\blacksquare}$)
$V_{\theta}=0.25$,
($\textcolor[rgb]{0.00,0.00,0.50}{\blacktriangleright}$)
$V_{\theta}=0.5$,
($\textcolor[rgb]{0.50,0.50,0.00}{\blacklozenge}$)
$V_{\theta}=1.0$,
($\textcolor[rgb]{1.00,0.50,0.00}{\blacktriangleleft}$)
$V_{\theta}=1.5$,
($\textcolor[rgb]{0.98,0.00,0.00}{\blacktriangle}$)
$V_{\theta}=2.5$. The solid line indicates the function
$v_{\theta}/V_{\theta}=\exp[-0.34(r-R_i)-0.0015(r-R_i)^2]$, and
the dashed one the function
$v_{\theta}/V_{\theta}=\exp[-0.21(r-R_i)-0.002(r-R_i)^2]$.
Geometry $R_{50}$.}}
\end{center}
\end{figure}

\begin{figure}[!htb]
\begin{center}
\includegraphics*[width=7cm]{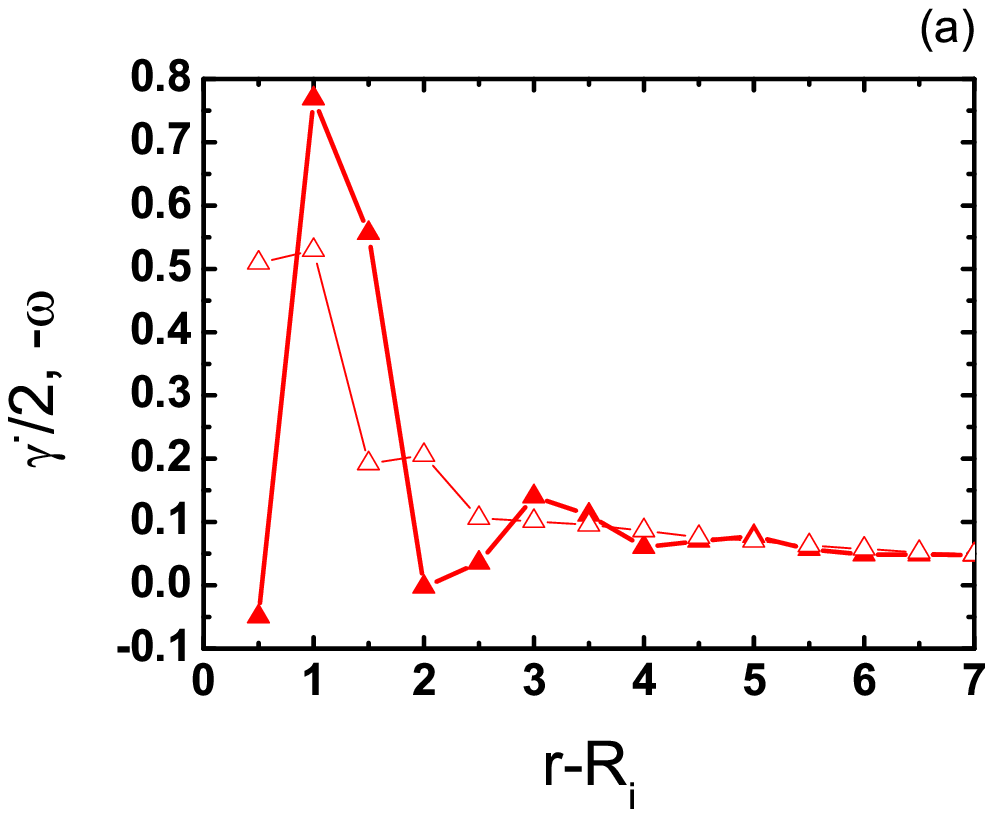}
\includegraphics*[width=7.25cm]{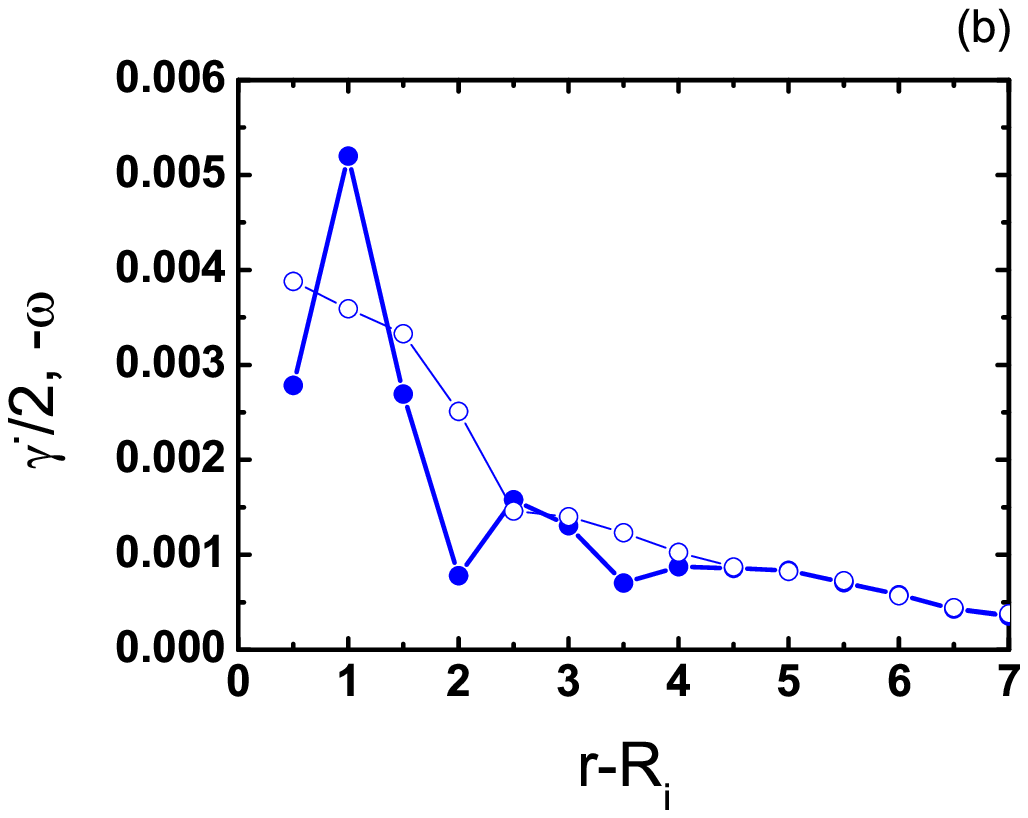}
\caption{\label{Fig6} \textit{(Color online) Influence of the
inner wall on the collapse between the average angular velocity
$\omega(r)$ (hollow symbols) and the shear rate $\dot \gamma (r)$
(full symbols) (a) $V_{\theta}=2.5$ and (b) $V_{\theta}=0.025$.
Geometry $R_{50}$.}}
\end{center}
\end{figure}

The shear localization near the inner wall is revealed by the
strong decrease of velocity profiles $v_{\theta}(r)$ shown on
Fig.~\ref{Fig4}. The decay appears to be nicely approximated by a
Gaussian function
$v_{\theta}/V_{\theta}=\exp[-a(r-R_i)-b(r-R_i)^2]$, as shown on
Fig.~\ref{Fig4}. We notice however that there is a sliding
velocity for the higher value of $V_{\theta}$ ($2.5$), which is
apparent in Fig.~\ref{Fig6}a. Previous studies in 2D
systems~\cite{Schollmann99, Howell99b, Latzel00, Losert00a,
Losert01, Latzel03} found an exponential shape, while a gaussian
decay was observed in three dimensional (3D) systems for non
spherical or polydispersed grains~\cite{Mueth00}. The agreement
between the measurement of the velocity profiles in 3D experiment
(using 3D MRI velocimetry in the bulk or CIV at the free
surface)~\cite{Gdr04} and 2D discrete simulations is
satisfactory~\cite{Dacruz04a}.

The normalization of $v_{\theta}(r)$ by shear velocity
$V_{\theta}$ allows to clearly visualize the influence of this
latter parameter on the velocity profiles. In the global
quasistatic regime ($V_{\theta}\leq 0.025$), there is no
influence, while for increasing $V_{\theta}$ above $0.025$, an
increase of the localization width is observed, consistently with
experimental observations~\cite{Losert01}.

The shear rate is equal to $\dot{\gamma}(r) = -r
\frac{\partial}{\partial r} (\frac{v_{\theta}(r)}{r})$. We denote
$\omega(r)$ the profile of the average angular velocity of the
grains. As previously reported in discrete simulations of granular
flows~\cite{Dacruz05}, the average angular velocity is equal to
half the local shear rate (or vorticity) $\omega(r) = -
\dot{\gamma}(r)/2$. Oscillations of the average angular velocity
are observed in the 3 or 4 first grain layers near the inner wall
(Fig.~\ref{Fig6}), as previously noticed by~\cite{Latzel00}. They
may be due to the frustration of the rotation of the flowing
grains in contact with the glued grains of the walls (which rotate
with angular velocity $\Omega$).

\subsection{Solid fraction}\label{sec:compa}

\begin{figure}[!htb]
\begin{center}
\includegraphics*[width=7cm]{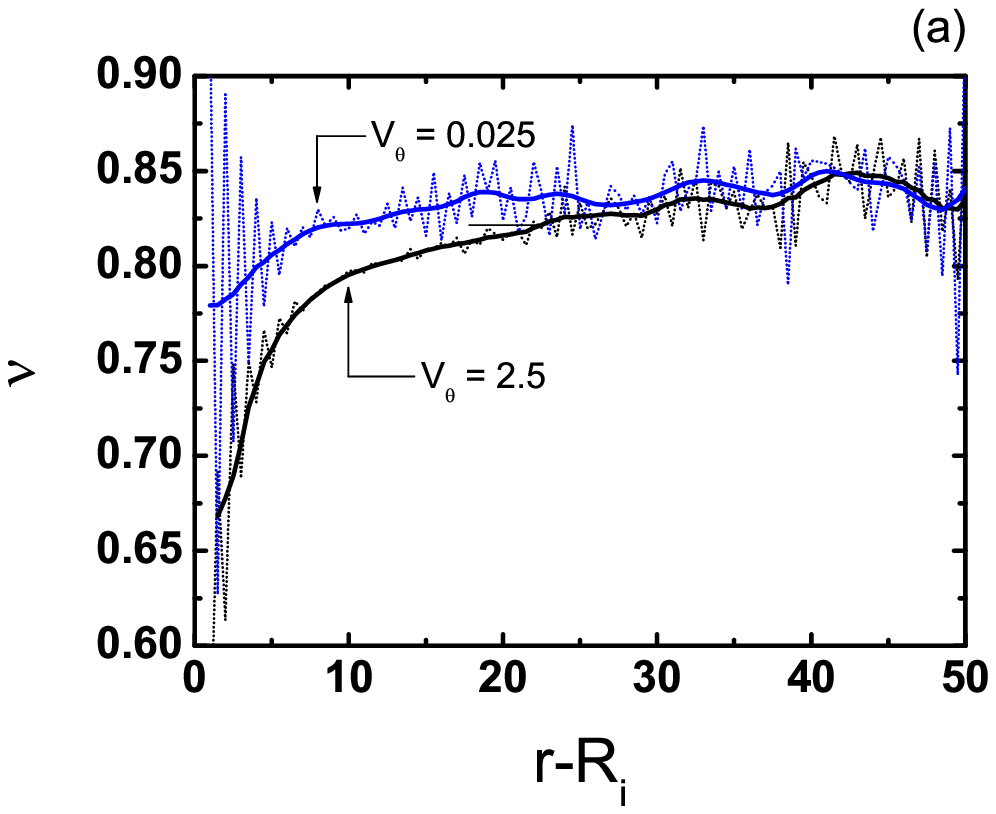}
\includegraphics*[width=7cm]{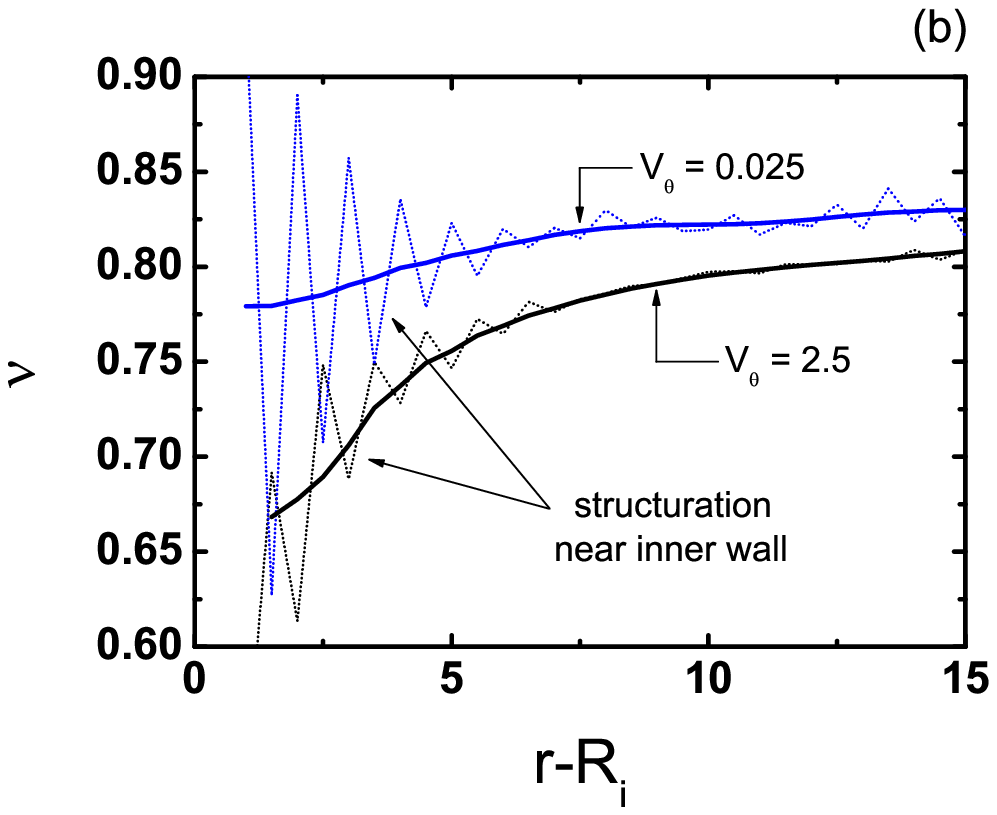}
\caption{\label{Fig7} \textit{(Color online) Influence of shear
velocity $V_{\theta}$ on the structuration near the inner wall.
Solid fraction profiles $\nu(r)$ (a) in the whole system and (b)
in the region close to the inner wall. The solid line is an
average over $3d$, while the dotted line is an average over
$0.5d$. Geometry $R_{50}$.}}
\end{center}
\end{figure}

Fig.~\ref{Fig7} shows the solid fraction profiles $\nu(r)$ for
$V_{\theta}=0.025$ and $2.5$. In the global quasistatic regime
($V_{\theta}\leq 0.025$), the profile becomes independent of
$V_{\theta}$, while a decrease of the solid fraction is observed
for increasing $V_{\theta}$. The material is significantly dilated
near the inner wall~\cite{Veje99, Schollmann99, Latzel00}, and is
structured in about $5$ layers close to the inner wall, with a
higher amplitude for low $V_{\theta}$. This was previously
observed in various shear geometries~\cite{Denniston99,
Schollmann99, Mueth00, Ertas01, Chevoir01a}. This structuration of
the granular material certainly affects the sliding of layers of
grains, with significant consequences on the mechanical behavior
near the wall. As previously reported~\cite{Latzel00, Latzel03},
independently of the influence of $V_{\theta}$, solid fraction
$\nu$ increases toward a value $\nu_{max}$ (close to $0.82$, the
solid fraction in the critical state for frictional disks with a
similar polydispersity~\cite{Radjai04}) away from the inner wall,
and remains close to its larger initial value $0.85$ in the region
where the material has not been sheared enough.

\section{Constitutive relations} \label{sec:behavior}

\begin{figure}[!htb]
\begin{center}
\includegraphics*[width=6cm]{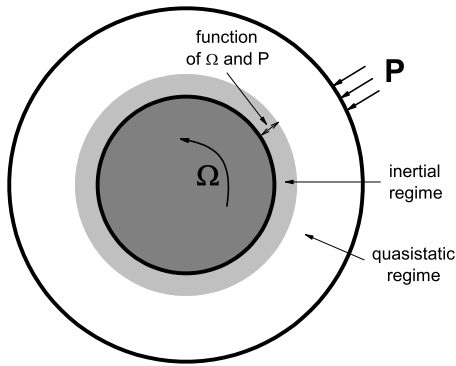}
\caption{\label{Fig12} \textit{Inertial and quasistatic zones.}}
\end{center}
\end{figure}

In Sec.~\ref{sec:localization} and App.~C, it has been shown that
shear velocity $V_{\theta}$, if small enough, no longer influences
the radial profiles of various quantities (see
Fig.~\ref{Fig3},~\ref{Fig4},~\ref{Fig8} and~\ref{Fig10}). Then,
the whole system is in the quasistatic regime. When $V_{\theta}$
increases, the shear rate $\dot{\gamma}$ increases in the whole
sample. Above a certain level of shear rate, inertial effects have
significant effect on the material behavior, which characterizes
the inertial regime. Considering the shear stress distribution in
the annular geometry and the decay of the velocity away from the
inner wall, we expect that the inertial zone begins at the inner
wall and that its thickness increases when $V_{\theta}$ increases
(Fig.~\ref{Fig12}).

In this section, we analyze the relations between different
dimensionless quantities in the inertial regime and how they are
affected by the transition to the quasistatic regime. We restrict
our analysis to large enough shear velocity ($V_{\theta}>0.025$),
so that a wide enough inertial zone exists close to the inner
wall.

\subsection{Inertial number and mechanical behavior}

Discrete simulations of homogeneous plane shear flows
~\cite{Dacruz05} have revealed that the constitutive law of dense
granular flows may be described through the dependency of the
effective friction $\mu^*$ (ratio of shear $\sigma$ to normal $P$
stresses) and of the solid fraction $\nu$ on the \emph{inertial
number} $I = \dot \gamma \sqrt{m/P}$ (a 2D equivalent of the
definition given in Sec.~\ref{sec:intro}), where all the
quantities are measured locally. The annular shear flows being
heterogeneous, we measure the relations between the local
quantities, $\nu(r)$, $\mu^*(r)=
\sigma_{r\theta}(r)/\sigma_{rr}(r)$ and $I(r) = \dot \gamma(r)
\sqrt{m/\sigma_{rr}(r)}$ (or $\dot \gamma(r)/
\sqrt{\sigma_{rr}(r)}$ in dimensionless unit). Each simulation
provides \emph{dynamic dilatancy} and \emph{friction laws} in a
range of inertial number. In the following, we try to analyze the
granular material as a continuum, consequently we do not take into
account the five first layers where wall structuration effects are
significant (see Fig.~\ref{Fig6} and Fig.~\ref{Fig7}).

\subsubsection{Dynamic friction law}

\begin{figure}[!htb]
\begin{center}
\includegraphics*[width=7.25cm]{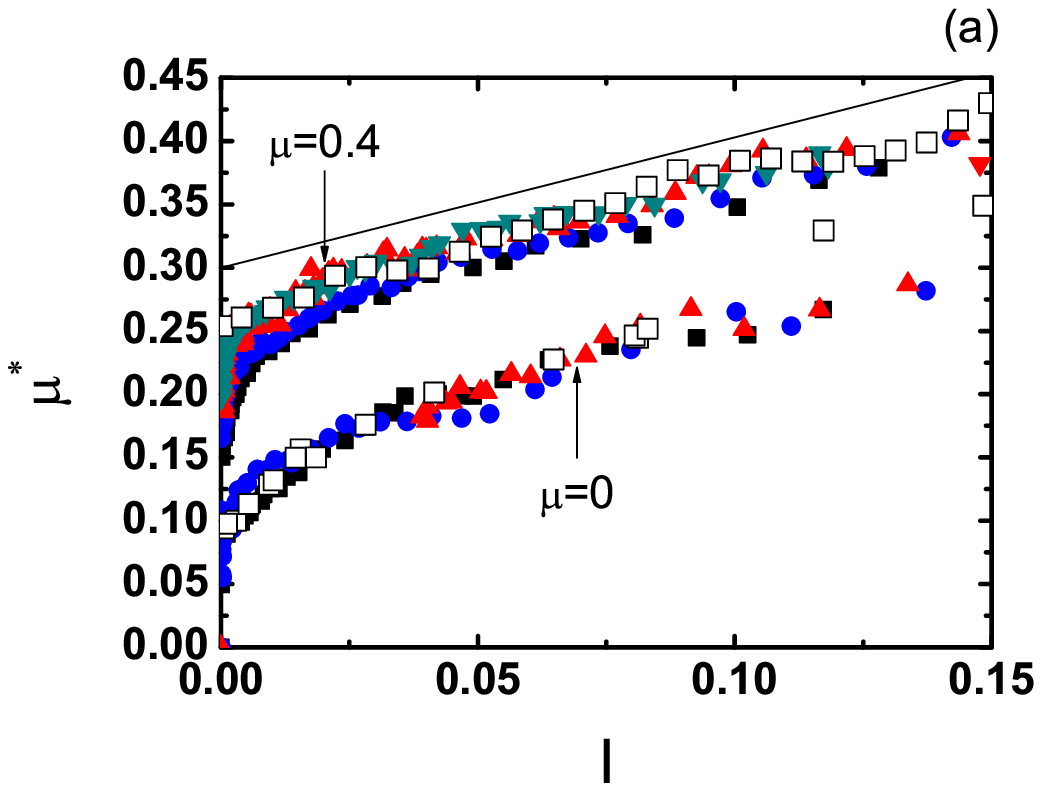}
\includegraphics*[width=7cm]{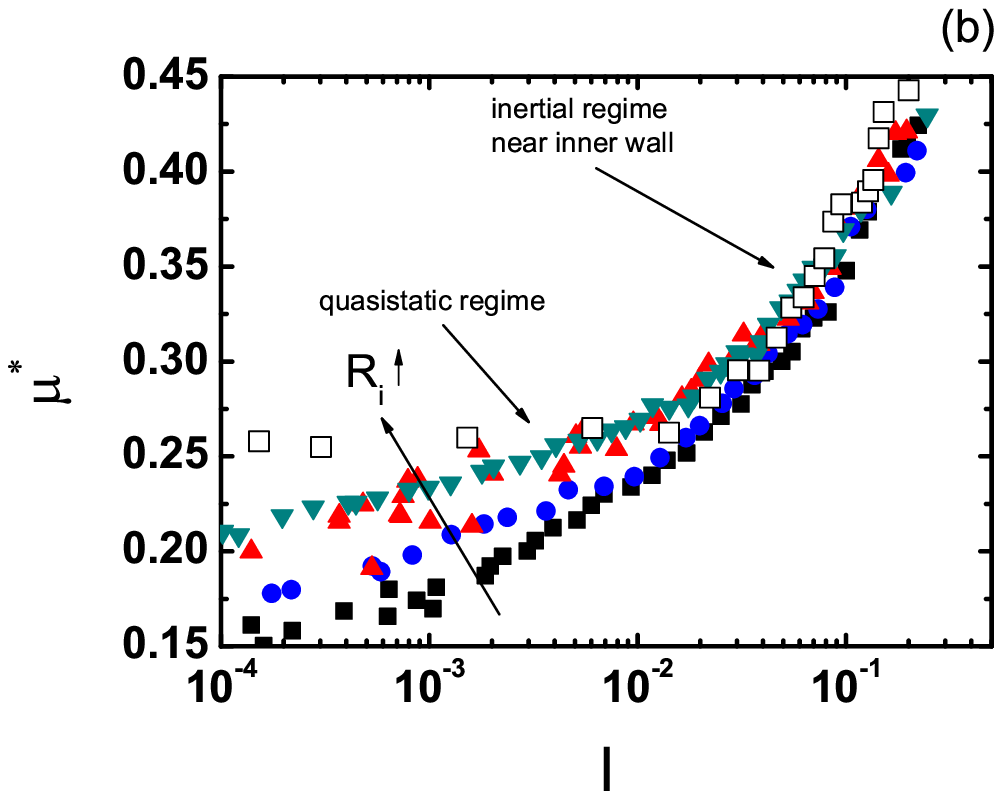}
\includegraphics*[width=7cm]{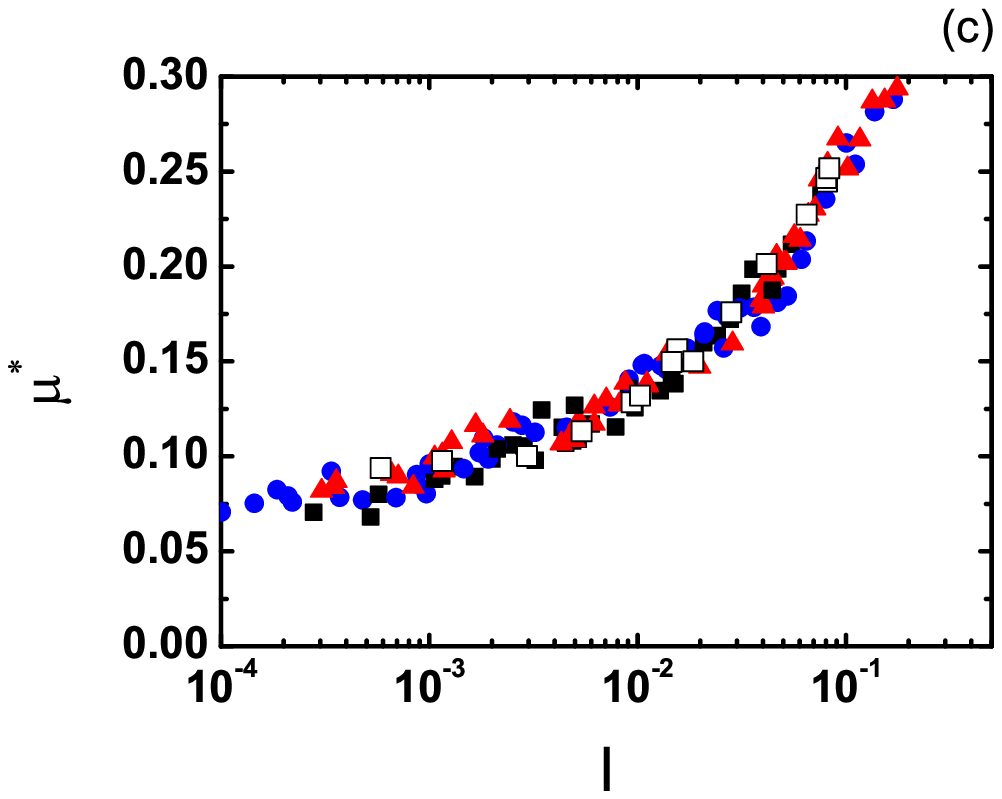}
\caption{\label{Fig13} \textit{(Color online) Dynamic friction law
(a) in linear scale (the solid line indicates a slope $\approx 1$)
for particle coefficient of friction $\mu=0$ and $\mu=0.4$.
Dynamic friction law in semi-logarithmic scale for (b) $\mu=0.4$
and (c) $\mu=0$. Different geometries: ($\blacksquare$) $R_{25}$,
($\textcolor[rgb]{0.00,0.00,1.00}{\bullet}$) $R_{50}$,
($\textcolor[rgb]{0.98,0.00,0.00}{\blacktriangle}$) $R_{100}$,
($\textcolor[rgb]{0.25,0.50,0.50}{\blacktriangledown}$) $R_{200}$.
$V_{\theta}=2.5$. Comparison with plane shear~\cite{Dacruz05}
($\square$).}}
\end{center}
\end{figure}

In the inertial regime, for $I \gtrsim 0.02$, $\mu^*$ increases
approximately linearly with $I$ and nearly independently of the
geometry (Fig.~\ref{Fig13}a):

\begin{equation}
    \label{eqn:mu(i)}
    \mu^{*}(I) \simeq \mu^*_{min} + bI,
\end{equation}

\noindent with $\mu^*_{min} \simeq 0.26$ and $b \simeq 1$. The
agreement with the dynamic friction law measured in the
homogeneous plane shear geometry is
excellent~\cite{Dacruz05,Rognon08a,Chevoir08b}. In contrast, for
lower values of $I$, a deviation from this linear relation is
observed, depending on the geometry (Fig.~\ref{Fig13}b). The
effective friction becomes smaller than $\mu^*_{min}$, and this
deviation increases as $R_i$ decreases, that is to say as the
stress gradient increases. Reciprocally, as $R_i$ increases, that
is to say as the stress distribution becomes more homogeneous, the
results of the annular geometry tend to the ones of the plane
shear geometry. This reveals that the simple relation between
effective friction $\mu^*$ and inertial number $I$ does not depend
on the stress distribution in the inertial regime, and is then
quite general (see ~\cite{Dacruz04a,Lois05} for flows down an
inclined plane), while it fails in the quasistatic regime. In
plane shear, $\mu^*_{min}$ may be considered as the \emph{internal
friction} in the critical state~\cite{Radjai04, Peyneau08}. This
is the maximum value of $\mu^*$ supported by the granular
material, before it starts to flow \emph{quasistatically}. With a
heterogeneous stress distribution, the granular material is able
to flow below this level.

We call $\lambda_{in}$ the width of the inertial zone. Using
Eqn.~\eqref{eqn:stressb} and \eqref{eqn:mu(i)}, we deduce that:

\begin{equation}
    \label{eqn:lambda}
    \lambda_{in}(V_{\theta}, R_i) = \left( \sqrt{S(V_{\theta}, R_i)/\mu^*_{min}}-1\right)
    R_i.
\end{equation}

\begin{figure}[!htb]
\begin{center}
\includegraphics*[width=6cm]{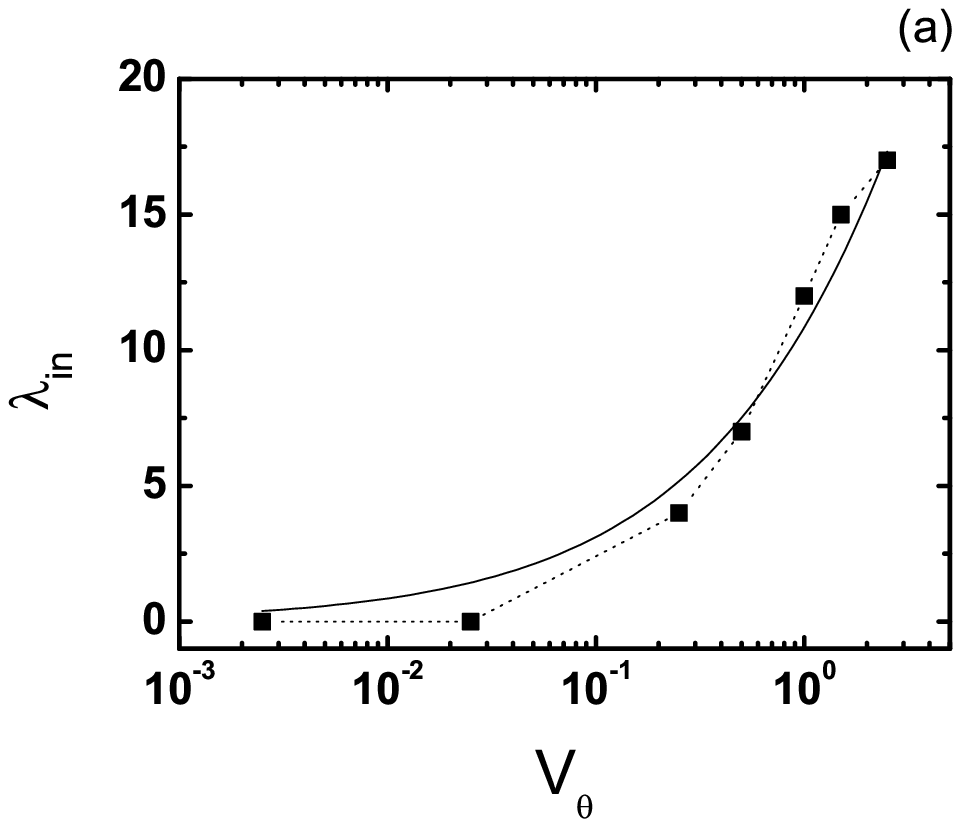}
\includegraphics*[width=6cm]{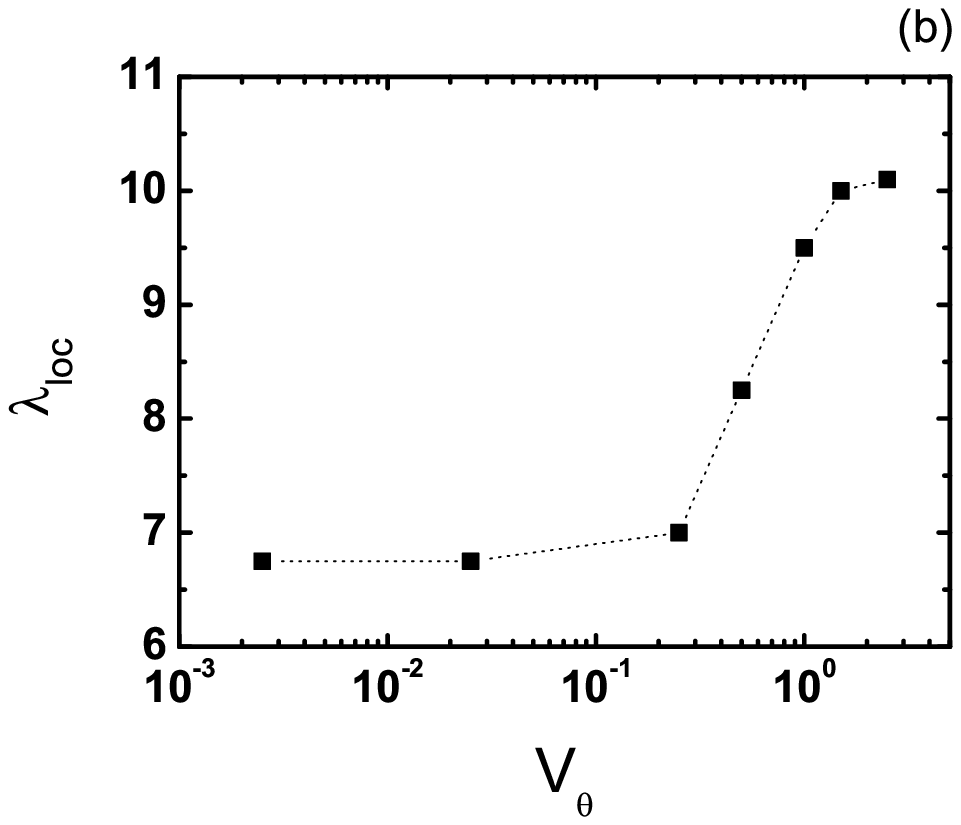}
\caption{\label{Fig14} \textit{(a) Width of the inertial zone
$\lambda_{in}$ as a function of $V_{\theta}$, as deduced from
Eqn.~\eqref{eqn:lambda} and Fig.~\ref{Fig3}a. The solid line
represents the function:
$\lambda_{in}=50(\sqrt{1+0.5V_{\theta}^{0.57}}-1)$. (b) Width of
the localization zone $\lambda_{loc}$ as a function of
$V_{\theta}$. Geometry $R_{50}$.}}
\end{center}
\end{figure}

We also conventionally define the width of the shear zone
$\lambda_{loc}$ through
$v_{\theta}(R_i+\lambda_{loc})=V_{\theta}/10$. Fig.~\ref{Fig14}a
and b shows $\lambda_{in}(V_{\theta})$ and
$\lambda_{loc}(V_{\theta})$ in geometry $R_{50}$. We notice that
$\lambda_{in}$ smoothly increases from zero with $V_{\theta}$,
while $\lambda_{loc}$ seems to saturate at a low value for low
$V_{\theta}$ (global quasistatic regime) and at a high value for
high $V_{\theta}$ (this is related to an apparent velocity
discontinuity near the wall, suggesting increasing collisional
effects in the first layers), with a sudden increase for
$V_{\theta}$ between $0.3$ and $1$. We notice that in the
experiment of~\cite{Losert01}, the shear zone invades the whole
gap for high enough $V_{\theta}$.

\begin{figure}[!htb]
\begin{center}
\includegraphics*[width=7cm]{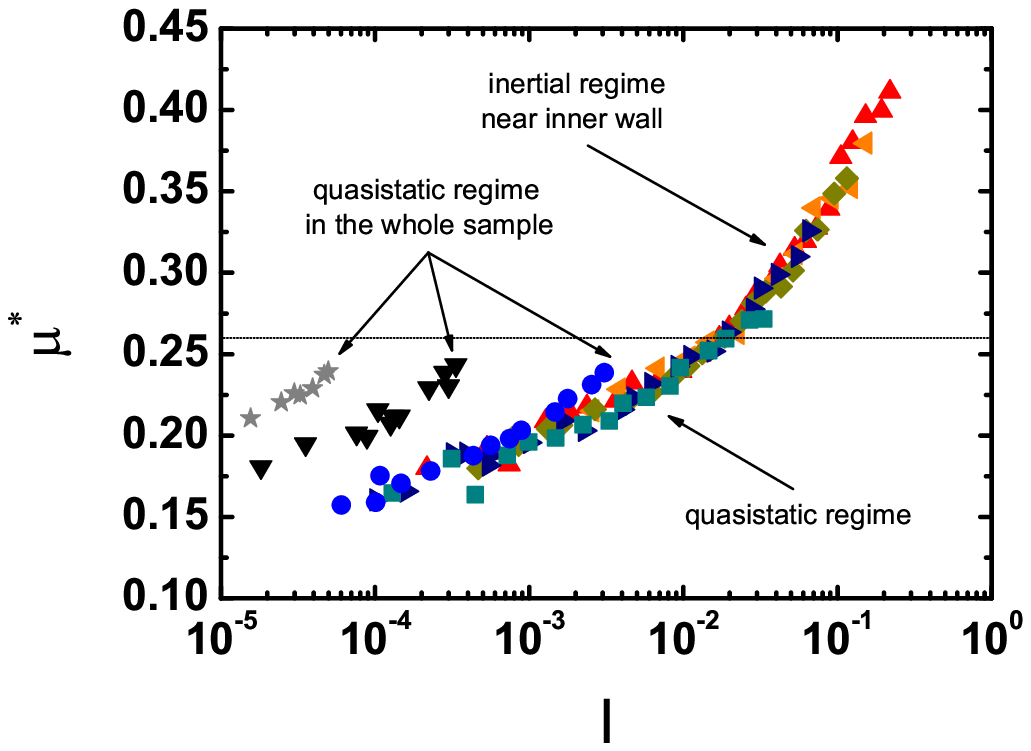}
\caption{\label{Fig15} \textit{(Color online) Effective friction
$\mu^*$ as a function of the inertial number $I$
($\textcolor[rgb]{0.5,0.5,0.5}{\bigstar}$) $V_{\theta}=0.00025$,
($\blacktriangledown$) $V_{\theta}=0.0025$,
($\textcolor[rgb]{0.00,0.00,1.00}{\bullet}$) $V_{\theta}=0.025$,
($\textcolor[rgb]{0.25,0.50,0.50}{\blacksquare}$)
$V_{\theta}=0.25$,
($\textcolor[rgb]{0.00,0.00,0.50}{\blacktriangleright}$)
$V_{\theta}=0.5$,
($\textcolor[rgb]{0.50,0.50,0.00}{\blacklozenge}$)
$V_{\theta}=1.0$,
($\textcolor[rgb]{1.00,0.50,0.00}{\blacktriangleleft}$)
$V_{\theta}=1.5$,
($\textcolor[rgb]{0.98,0.00,0.00}{\blacktriangle}$)
$V_{\theta}=2.5$. The solid line corresponds to $\mu^* = 0.26$.
Geometry $R_{50}$.}}
\end{center}
\end{figure}

In a given geometry, for a small enough shear velocity
$V_{\theta}$, the inertial zone disappears, and the whole system
is in the quasistatic regime. Fig.~\ref{Fig15} then shows again
that the effective friction $\mu^*$ is no more a function of $I$.

\subsubsection{Dynamic dilatancy law}

\begin{figure}[!htb]
\begin{center}
\includegraphics*[width=7cm]{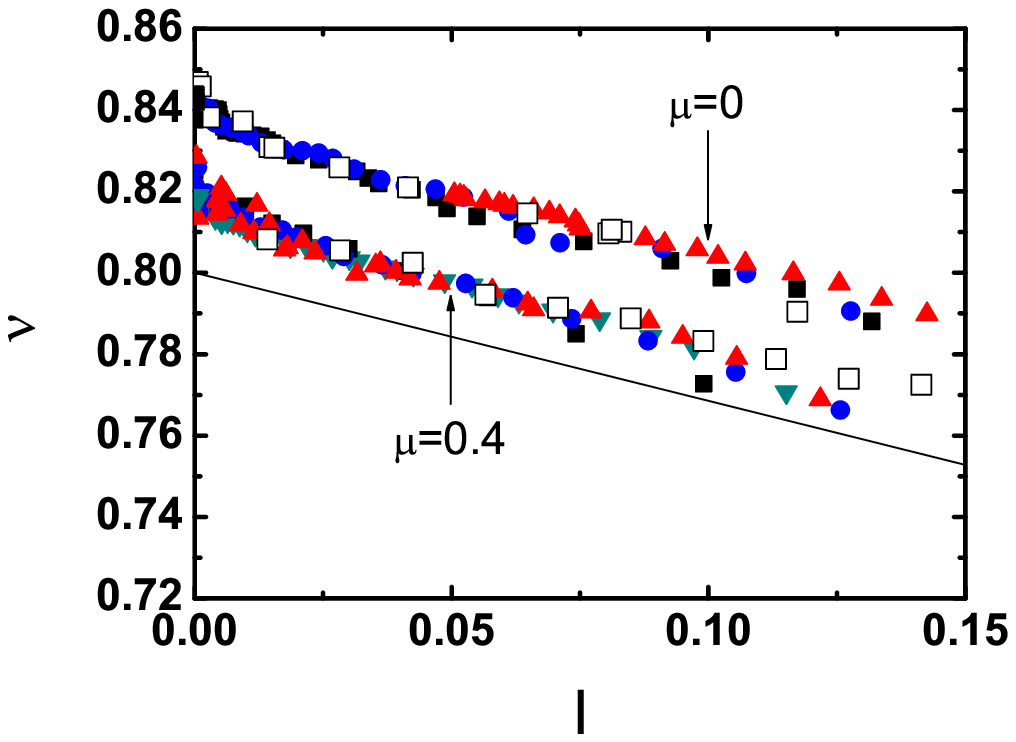}
\caption{\label{Fig16} \textit{(Color online) Dynamic dilatancy
law (the solid line indicate a slope $\approx -0.37$) for
different geometries: ($\blacksquare$) $R_{25}$,
($\textcolor[rgb]{0.00,0.00,1.00}{\bullet}$) $R_{50}$,
($\textcolor[rgb]{0.98,0.00,0.00}{\blacktriangle}$) $R_{100}$,
($\textcolor[rgb]{0.25,0.50,0.50}{\blacktriangledown}$) $R_{200}$.
Comparison with plane shear~\cite{Dacruz05} ($\square$).}}
\end{center}
\end{figure}

\begin{figure}[!htb]
\begin{center}
\includegraphics*[width=7cm]{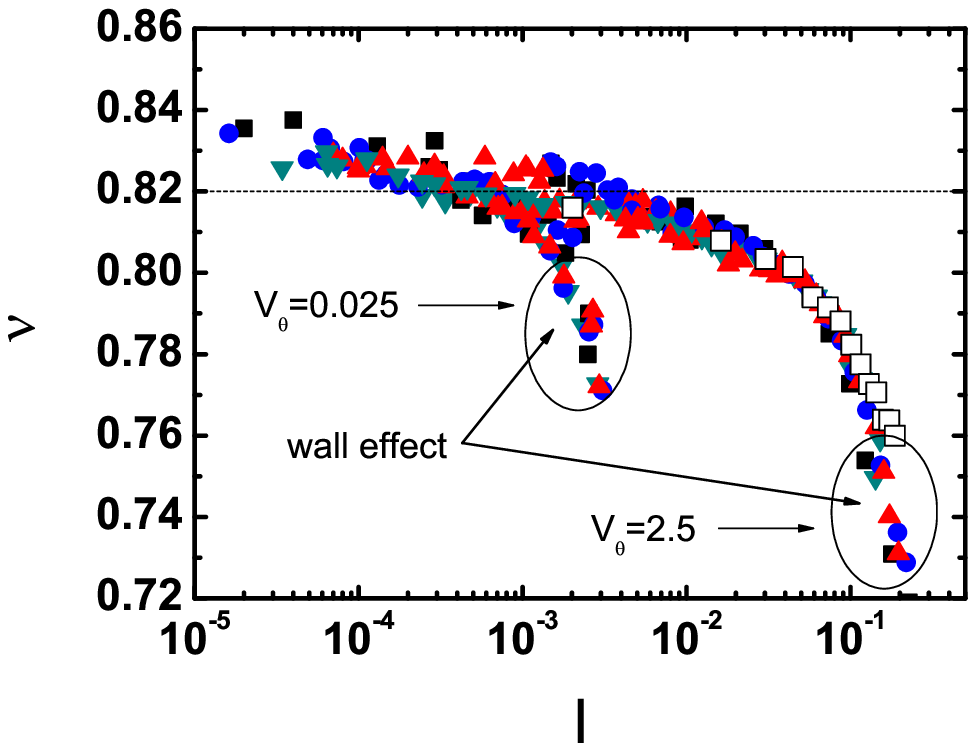}
\caption{\label{Fig17} \textit{(Color online) Dynamic dilatancy
law in semi-logarithmic scale. Different geometries:
($\blacksquare$) $R_{25}$,
($\textcolor[rgb]{0.00,0.00,1.00}{\bullet}$) $R_{50}$,
($\textcolor[rgb]{0.98,0.00,0.00}{\blacktriangle}$) $R_{100}$,
($\textcolor[rgb]{0.25,0.50,0.50}{\blacktriangledown}$) $R_{200}$.
Comparison with plane shear~\cite{Dacruz05} ($\square$). The
dashed line corresponds to $\nu_{max} = 0.82$.}}
\end{center}
\end{figure}

We observe a linear decrease of solid fraction $\nu$ as a function
of inertial number $I$, independently of the geometry in the
inertial regime (Fig.~\ref{Fig16} and~\ref{Fig17}), and $\nu$
tends to a maximum value $\nu_{max}$, which identifies to the
solid fraction in the critical state. We can then write the
dynamic dilatancy law:

\begin{equation}
    \label{eqn:nu(i)}
    \nu(I) \simeq \nu_{max} - aI,
\end{equation}

\noindent with $\nu_{max} \simeq 0.82$ and $a \simeq 0.37$. The
agreement with the dynamic dilatancy law measured in the
homogeneous plane shear geometry is
excellent~\cite{Dacruz05,Rognon08a}. However, far from the walls,
in the region where the material is less deformed and so remains
in its initial dense state, higher values of $\nu$ are observed.
Fig.~\ref{Fig17} also indicates that the inner wall induces
further dilation.

\subsubsection{Frictionless grains}

As shown in Fig.~\ref{Fig13}a and Fig.~\ref{Fig16}, the
microscopic friction coefficient $\mu$ has a significant influence
on the constitutive law parameters. Those figures also reveal good
agreement with homogeneous shear simulations~\cite{Dacruz05}. The
solid fraction remains a linearly decreasing function of $I$ (with
a fast change in the quasistatic limit). The slope $a$ is not
affected, while $\nu_{max}$ increases to $\simeq 0.85$. The
dynamic friction law keeps the same tendency but is shifted toward
smaller values of friction. The linear approximation with
$\mu^*_{min} \simeq 0.11$ (Eqn.~\eqref{eqn:mu(i)}) fails for $I
\le 0.01$. We notice that the range of validity of the dynamic
friction law is much larger than for frictional grains, and that
it does not seem to depend on the geometry.

Those differences are likely related to some peculiarities of
assemblies of frictionless grains~\cite{Peyneau08}. The
quasistatic limit, in such materials, is only approached for much
smaller values of $I$ than in the frictional case, and
$\mu^*_{min}$ is itself considerably lower. As a consequence on
may expect a wider inertial zone. Moreover, as the critical solid
fraction coincides with the random close packing
value~\cite{Peyneau08}, no solid-like region of the system can be
prevented from flowing because of its density.

\subsubsection{Comparison with previous studies}

The validity of the constitutive law, once suitably generalized to
three dimensions, was successfully tested in flows down a heap
between lateral walls~\cite{Jop05b,Jop06}. In that case the
velocity field, as deduced from numerical computations in which
the viscoplastic law was implemented, exhibits a more complex
three-dimensional structure. Predicted velocities at the free
surface agreed closely with experimental results.

Thus, the applicability of the constitutive law as a relation
between local values of non-uniform strain rate and stress fields,
which we just established in 2D annular shear flow, was previously
checked in the 3D case of a laterally confined gravity-driven
flow. The validity of such an approach should be restricted to
situations in which the characteristic length for stress or strain
rate variations, say $l$, is significantly larger than the grain
size. In annular shear, one has $l=R_i$, whereas the finite width
$w$ of the channel was found in~\cite{Jop05b,Jop06} to control the
gradients, $l=w$. As $R_i$, in units of grain diameters, varies
between 25 and 200 here, while the interval of $w$ extends between
16.5 and 500 in~\cite{Jop05b,Jop06}, similar levels of
heterogeneity are explored.

\subsection{Internal variables}

We now discuss how internal variables, which profiles are
discussed in App.~C, scale with the inertial number $I$, revealing
local state laws, consistent with the one measured in homogeneous
shear flows.

\begin{figure}[!htb]
        \begin{center}
            \includegraphics*[width=7cm]{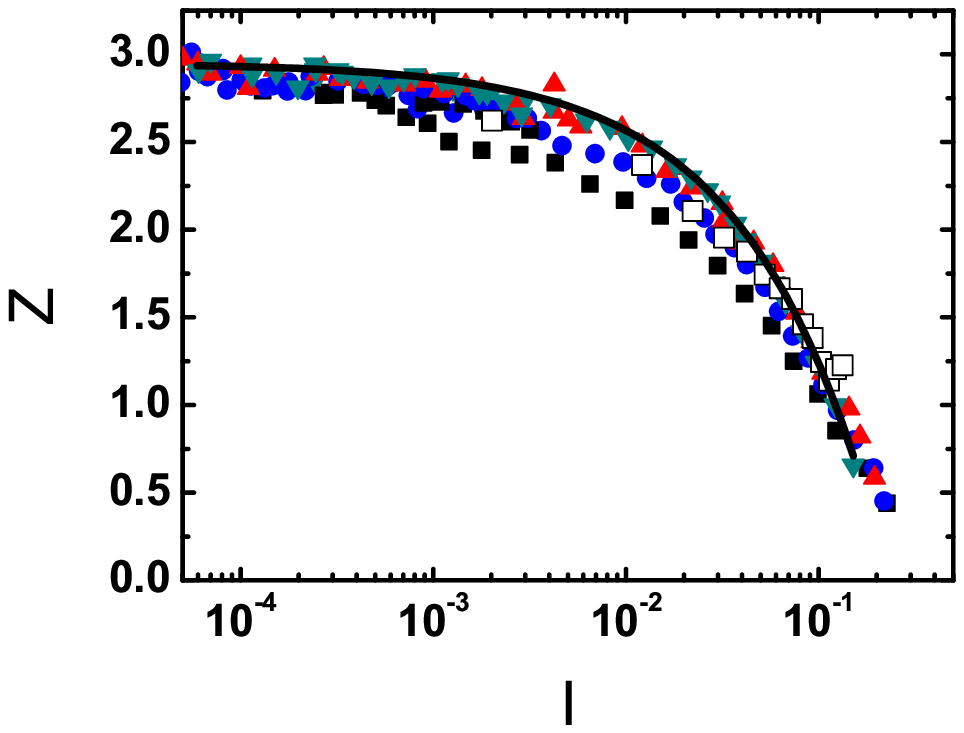}
            \caption{\label{Fig19}
            \textit{(Color online) Coordination number $Z$ as a function of the
            inertial number $I$ (the solid line represents the function
            $Z=2.95-7.65I^{0.65}$) for different geometries. ($\blacksquare$) $R_{25}$,
($\textcolor[rgb]{0.00,0.00,1.00}{\bullet}$) $R_{50}$,
($\textcolor[rgb]{0.98,0.00,0.00}{\blacktriangle}$) $R_{100}$,
($\textcolor[rgb]{0.25,0.50,0.50}{\blacktriangledown}$) $R_{200}$,
($\square$) plane shear~\cite{Dacruz05}. $V_{\theta}=2.5$.}}
        \end{center}
    \end{figure}

We observe a relation like $Z=Z_{max}-eI^f$ (with $Z_{max} \approx
3$) between coordination number $Z$ and $I$ on Fig.~\ref{Fig19},
nearly independent of the geometry.

\begin{figure}[!htb]
        \begin{center}
            \includegraphics*[width=7cm]{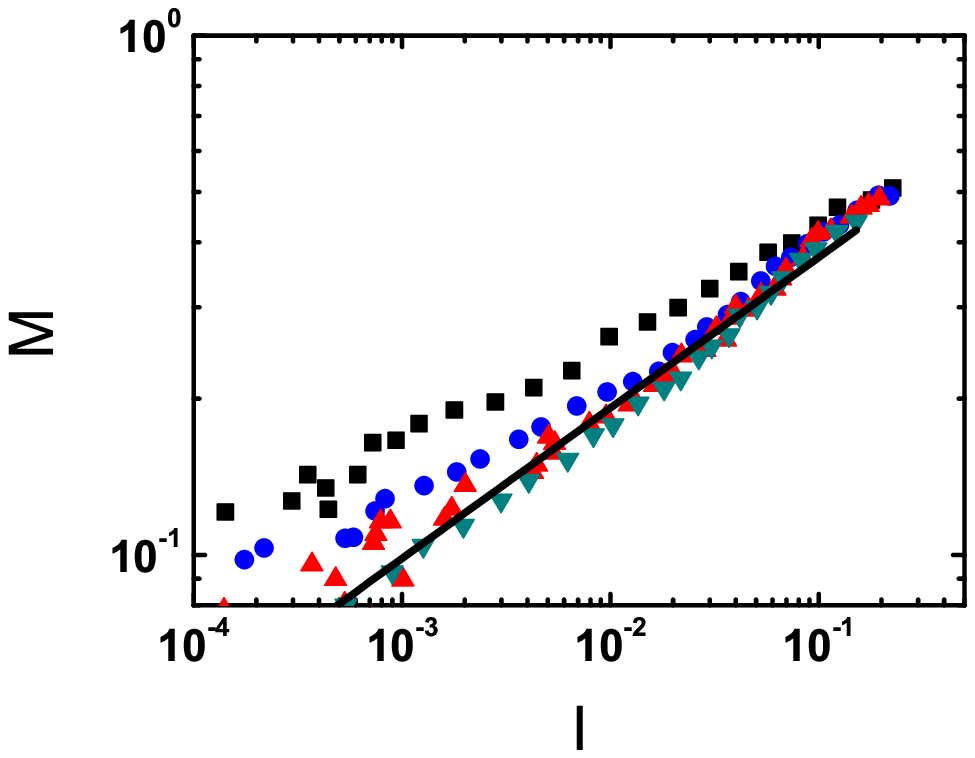}
            \caption{\label{Fig20}
            \textit{(Color online) Mobilization of friction $M$ as a function
of the inertial number $I$ (the solid line represents the function
$M=0.73I^{0.29}$ for different geometries: ($\blacksquare$)
$R_{25}$, ($\textcolor[rgb]{0.00,0.00,1.00}{\bullet}$) $R_{50}$,
($\textcolor[rgb]{0.98,0.00,0.00}{\blacktriangle}$) $R_{100}$,
($\textcolor[rgb]{0.25,0.50,0.50}{\blacktriangledown}$) $R_{200}$.
$V_{\theta}=2.5$.}}
        \end{center}
    \end{figure}

We do not observe a general relation between the mobilization of
friction $M$ and $I$, but an asymptotic convergence for growing
$R_i$ toward a relation $M \approx gI^h$ (Fig.~\ref{Fig20}). For
this quantity, there is no satisfactory agreement with the
homogeneous shear case.

\begin{figure}[!htb]
\begin{center}
\includegraphics*[width=7cm]{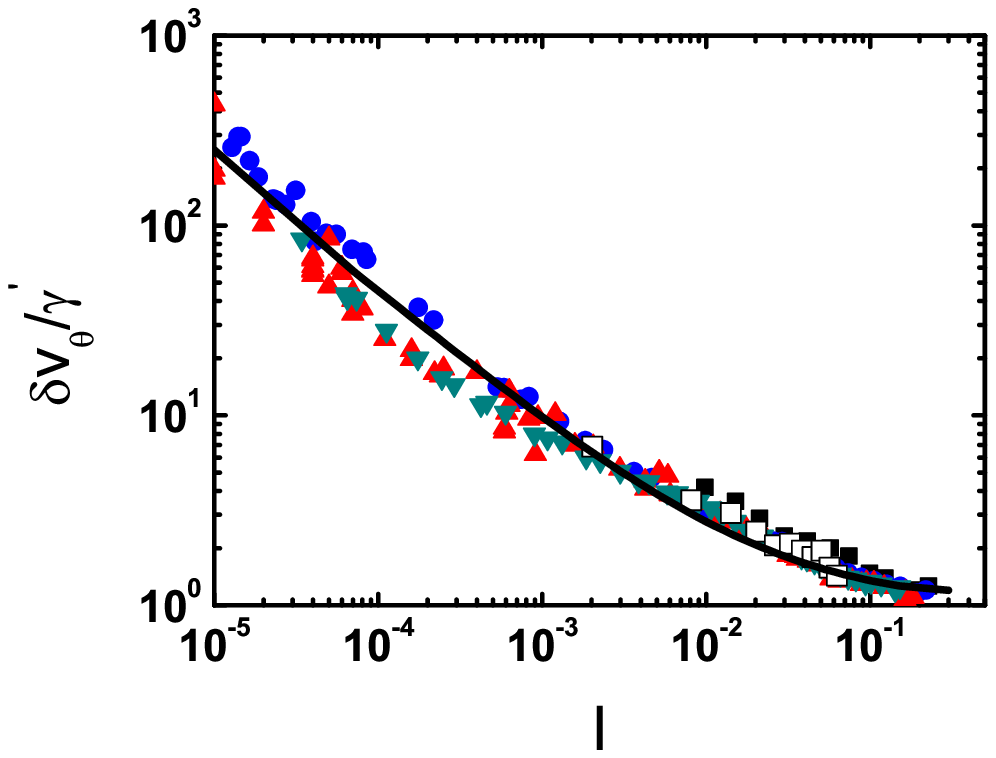}
\caption{\label{Fig21} \textit{(Color online) Relative
fluctuations as a function of the inertial number $I$ (the solid
line represents the function $\delta
v_{\theta}/(\dot{\gamma}d)=1+0.07I^{-0.7}$). ($\blacksquare$)
$R_{25}$, ($\textcolor[rgb]{0.00,0.00,1.00}{\bullet}$) $R_{50}$,
($\textcolor[rgb]{0.98,0.00,0.00}{\blacktriangle}$) $R_{100}$,
($\textcolor[rgb]{0.25,0.50,0.50}{\blacktriangledown}$) $R_{200}$,
$\square$ plane shear~\cite{Dacruz05}. $V_{\theta}=2.5$.}}
\end{center}
\end{figure}

We analyse the fluctuations of orthoradial velocity $\delta
v_{\theta}$ normalized by the natural scale $\dot{\gamma}$ as a
function of $I$ (Fig.~\ref{Fig21}). In the quasistatic regime, the
development of collective and intermittent motions
(see~\cite{Mueth03,Chambon03a} in annular shear and
~\cite{Mills08} for a recent review) explain the increase of these
relative fluctuations. For higher values of the inertial number
$I$, we observe that $\delta v_{\theta}/\dot{\gamma} \rightarrow
1$. On the whole, we propose to describe the dependency by the
equation $\delta v_{\theta}/\dot{\gamma}=1+cI^{-d}$
(Fig.~\ref{Fig21}). Experimental results~\cite{Mueth00,
Bocquet02b} show that $\delta v \varpropto \dot{\gamma}^{0.4}$.
Dividing this relation by $\dot\gamma$, we get an exponent equal
to $-0.6$, close to exponent $d = -0.7$, deduced from the previous
fit.

\section{Consequences for the shear localization and the
macroscopic behavior} \label{sec:prediction}

Using the constitutive law established in
Sec.~\ref{sec:behavior}, we now show that it is possible to
understand some observations described in
Sec.~\ref{sec:localization}.

\begin{figure}[!htb]
\begin{center}
\includegraphics*[width=7.5cm]{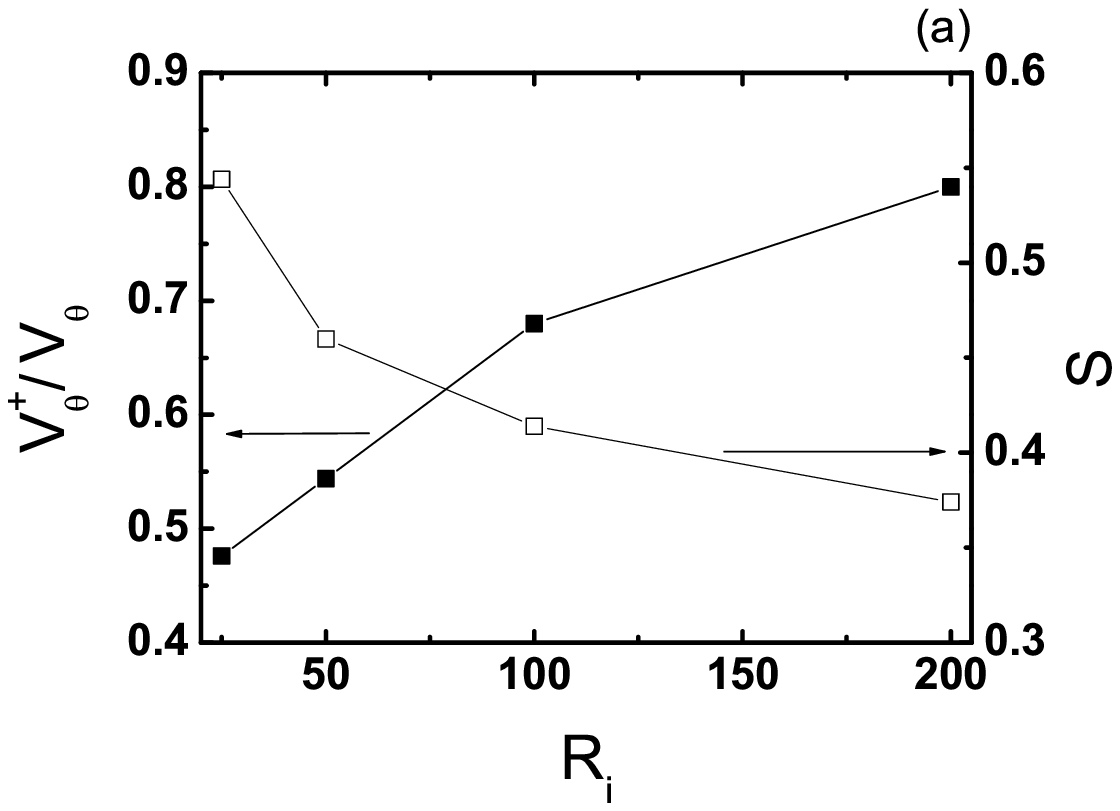}
\includegraphics*[width=7cm]{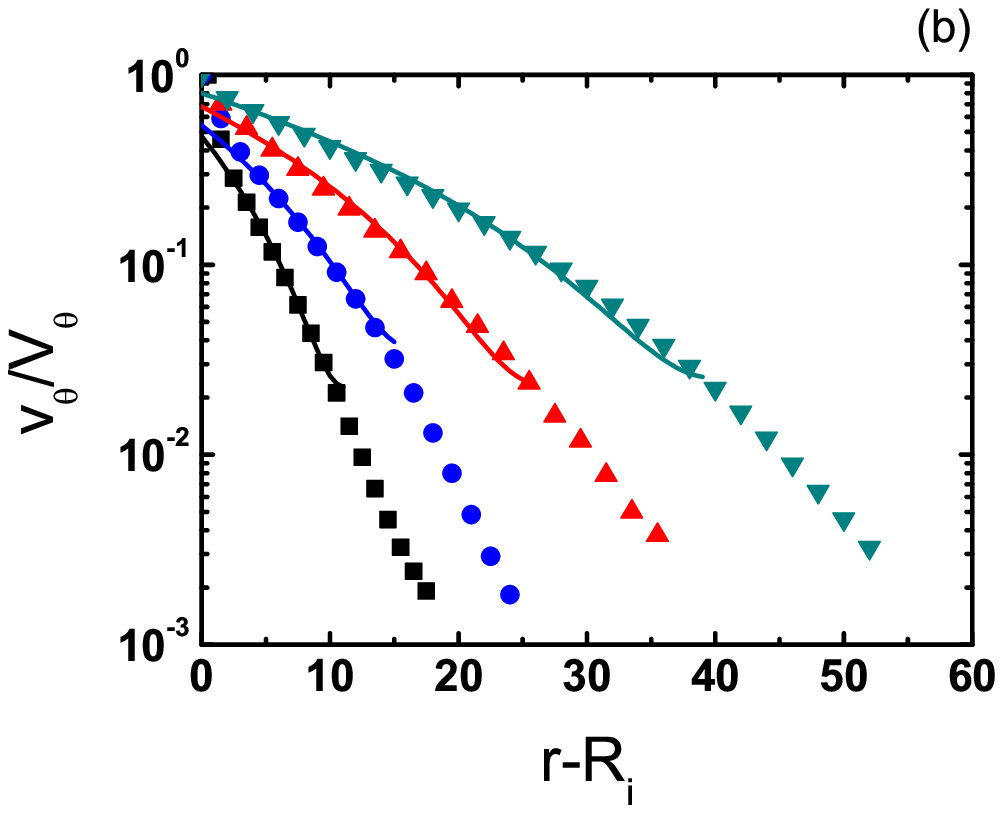}
\caption{\label{Fig22} \textit{(Color online) (a) Influence of the
geometry on ($\blacksquare$) $V^+_{\theta}/V_{\theta}$ (fit) and
($\square$) $S$ (measurement) ($V_{\theta}=2.5$). (b) Velocity
profiles: comparison between the measurements ($\blacksquare$)
$R_{25}$, ($\textcolor[rgb]{0.00,0.00,1.00}{\bullet}$) $R_{50}$,
($\textcolor[rgb]{0.98,0.00,0.00}{\blacktriangle}$) $R_{100}$,
($\textcolor[rgb]{0.25,0.50,0.50}{\blacktriangledown}$) $R_{200}$
and the prediction of Eqn.~\eqref{eqn:fitv} (solid lines). The
velocity profiles are limited to the steady zone $R_i \to R_i +
R_{steady}$ ($V_{\theta}=2.5$).}}
\end{center}
\end{figure}

Still using dimensionless units, since the pressure $P$ is
constant in the system and the shear stress is given by
Eqn.~\eqref{eqn:stressb}, the dynamic friction law
Eqn.~\eqref{eqn:mu(i)} provides the following equation for the
velocity profile $v_{\theta}(r)$:

\begin{equation}
\frac{\partial}{\partial r} (\frac{v_{\theta}(r)}{r})  =
\frac{\mu^*_{min}}{br} -\frac{S R_i^2}{br^3},
\end{equation}

\noindent where the shear stress at the inner wall $S$ depends
both on $V_{\theta}$ and on $R_i$ (see Sec.~\ref{sec:stress}). As
shown from the measurements drawn in Fig.~\ref{Fig22}b, for a
large value of $V_{\theta}$, $S$ is high in small geometries and
strongly decreases as $R_i$ increases. We now integrate this
relation over the range of validity of the dynamic friction law,
this is to say in the inertial zone $R_i \to R_{in} = R_i
+\lambda_{in}$, from which we get:

\begin{equation}
v_{\theta}(r)  = \frac{S R_i^2}{2br} + \frac{\mu^*_{min}}{b} r \ln
(r) + cr.
\end{equation}

\noindent The constant $c$ is determined by the value of the
velocity at the inner wall, called $V^+_{\theta}$, which is
smaller than $V_{\theta}$, revealing some sliding at the wall as
previously noticed (see Sec.~\ref{sec:velocity}).

\begin{equation}
V^+_{\theta}  = \frac{S R_i}{2b} + \frac{\mu^*_{min}}{b} R_i \ln
(R_i) + cR_i.
\end{equation}

On the whole, the velocity profile is equal to:

\begin{equation}
\label{eqn:fitv} v_{\theta}(r)=V^+_{\theta} \frac{r}{R_i}+r
    \left[\frac{S}{2b}\left(\left(\frac{R_i}{r}\right)^2-1\right)+
    \frac{\mu^*_{min}}{b} \ln \left(\frac{r}{R_i}\right)\right].
\end{equation}

An absolute measurement of $V^+_{\theta}$ happens to be difficult,
considering the wall effect that disturbs the material behavior in
a layer of a few grains near the inner wall (like in
Fig.~\ref{Fig6} for the shear rate $\dot{\gamma}$). Consequently,
we obtain this quantity from a fit, and a comparison with the
measured velocity profiles is shown on Fig.~\ref{Fig22}b. The
agreement is excellent, suggesting once more the validity of the
dynamic friction law. The sliding increases when $V_{\theta}$
increases and, as shown in Fig.~\ref{Fig22}b, increases when $R_i$
decreases.

We now try to predict the $S(V_{\theta})$ relation, which was
measured and fitted in Fig.~\ref{Fig3}b. It is clear that in the
global quasistatic limit, as $V_{\theta} \to 0$, $S \to
\mu^*_{min}$. We now write a boundary condition at the limit of
the inertial zone $R_{in}$. Having used the dynamic friction law
Eqn.~\eqref{eqn:mu(i)}, we necessarily have $\dot \gamma (R_{in})
= 0$, as appears in the fitted curve in Fig~\ref{Fig22}b. Then,
beyond $R_{in}$, if this dynamic friction law was still valid ,
$\dot \gamma (r)$ would be strictly equal to zero, so that
$v_{\theta}(r)$ would be equal to $Cr$, with a constant $C$. The
sole possibility is $C=0$ since the velocity must be equal to zero
at the outer wall. We conclude that $v_{in}=v_{\theta}(R_{in})=0$.
This conclusion is wrong, as it is clear in Fig~\ref{Fig22}b, and
has already been  discussed: the dynamic friction law fails in the
quasistatic regime, and we shall come back to this point just
after the discussion of the $S(V_{\theta})$ relation. The previous
assumption writes:

\begin{equation}
0 =V^+_{\theta} \frac{R_{in}}{R_i}+R_{in}
    \left[\frac{S}{2b}\left(\left(\frac{R_i}{R_{in}}\right)^2-1\right)+
    \frac{\mu^*_{min}}{b} \ln \left(\frac{R_{in}}{R_i}\right)\right].
\end{equation}

\noindent Since Eqn.~\eqref{eqn:lambda} is equivalent to
$R_{in}/R_i=\sqrt{S/\mu^*_{min}}$, we get the following implicit
$S(V^+_{\theta})$ relation:

\begin{equation}\label{eqn:SVtheta}
\frac{S}{\mu^*_{min}}-1-\ln\left(\frac{S}{\mu^*_{min}}\right)=
\frac{2bV^+_{\theta}}{\mu^*_{min} R_i}.
\end{equation}

\begin{figure}[!htb]
\begin{center}
\includegraphics*[width=7cm]{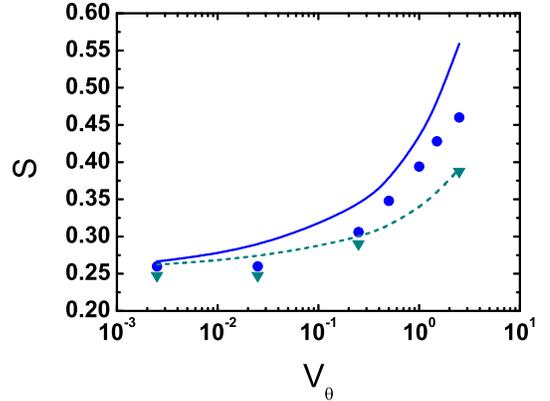}
\caption{\label{Fig23} \textit{(Color online) Shear stress at the
wall $S$ as a function of the wall velocity $V_{\theta}$.
Comparison between the measurements:
($\textcolor[rgb]{0.00,0.00,1.00}{\bullet}$) $R_{50}$ and
($\textcolor[rgb]{0.25,0.50,0.50}{\blacktriangledown}$) $R_{200}$
and the predictions of Eqn.~\eqref{eqn:SVtheta}. The solid and
dashed line respectively indicate the results for $R_{50}$ and
$R_{200}.$}}
\end{center}
\end{figure}

\noindent For simplicity, we take $V^+_{\theta}=V_{\theta}$ in the
comparison with the measurements, drawn in Fig~\ref{Fig23} for two
geometries. The agreement is very satisfactory considering the
previous simplifying assumptions. With the increase of $R_i$, the
difference between $V^+_{\theta}$ and $V_{\theta}$ decreases,
explaining the better results for $R_{200}$. For small
$V_{\theta}$, we write $S(V_{\theta}) = \mu^*_{min}
(1+f(V_{\theta}))$. A simple development gives $f \simeq
\sqrt{\frac{2b}{\mu^*_{min}R_i}} \sqrt{V_{\theta}}$. For $R_i =
50$, $f \simeq 0.55 \sqrt{V_{\theta}}$, which is close to the fit
$f \simeq 0.5 V_{\theta}^{0.57}$ used in Fig.~\ref{Fig3}b.
According to this analysis, $S$ becomes proportional to
$V_{\theta}$ for much larger values, not usually accessible.

\begin{figure}[!htb]
\begin{center}
\includegraphics*[width=7cm]{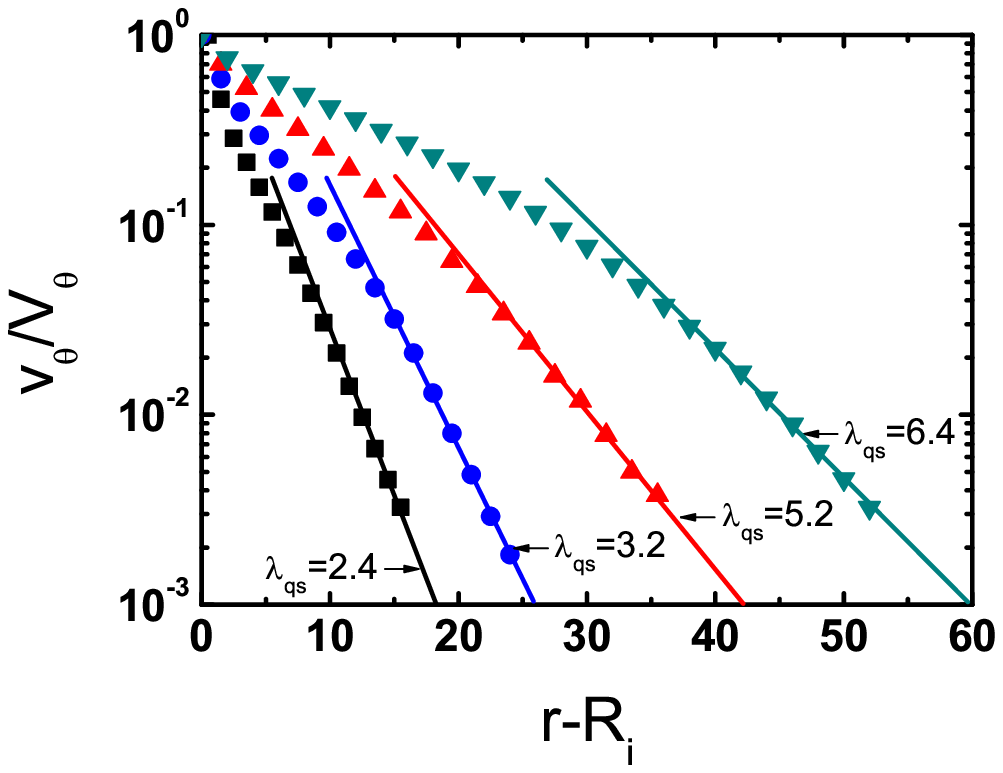}
\caption{\label{Fig24} \textit{(Color online) Characteristic
length $\lambda_{qs}$ obtained from the velocity profiles in the
quasistatic zone. Different geometries: ($\blacksquare$)
$R_{25}$, ($\textcolor[rgb]{0.00,0.00,1.00}{\bullet}$) $R_{50}$,
($\textcolor[rgb]{0.98,0.00,0.00}{\blacktriangle}$) $R_{100}$,
($\textcolor[rgb]{0.25,0.50,0.50}{\blacktriangledown}$)
$R_{200}$. $V_{\theta}=2.5$.}}
\end{center}
\end{figure}

We now come back to the limit of the dynamic friction law,
Eqn.~\eqref{eqn:mu(i)}, in the quasistatic limit, as shown in
Fig.~\ref{Fig13}b. A large portion of the velocity profile in the
steady quasistatic regime is shown in Fig.~\ref{Fig22}b. As a
first approximation, the velocity can be considered exponential in
this region, so that we write:

\begin{equation}
v_{\theta}(r)=v_{in} \exp -\left(\frac{r-R_{in}}{\lambda_{qs}}
\right),
\end{equation}

\noindent with $\lambda_{qs}$ the characteristic length in the
quasistatic region, measured in Fig.~\ref{Fig24} (which are
slightly larger than the one estimated for very small
$V_{\theta}$, that is to say when the quasistatic zone invades
the system), and $v_{in}$ is not equal to zero contrarily to the
previous simple approximation but, using Eqn.~\eqref{eqn:fitv}
to:

\begin{equation}
v_{in} = \sqrt{\frac{S}{\mu^*_{min}}} \left( V^+_{\theta} +
\frac{\mu^*_{min}R_i}{2b} \left( \ln
\left(\frac{S}{\mu^*_{min}}\right) - \frac{S}{\mu^*_{min}} + 1
\right) \right).
\end{equation}

\noindent Still using dimensionless units, since the pressure $P$
is constant in the system, we deduce that, for $r \ge R_{in}$, the
inertial number is equal to:

\begin{equation}
I(r) = \left( \frac{1}{\lambda_{qs}} + \frac{1}{r}\right) v_{in}
\exp -\left(\frac{r-R_{in}}{\lambda_{qs}}\right).
\end{equation}

\noindent Since $R_{in} \gg \lambda_{qs}$, we may write:

\begin{equation}
I(\mu^*) \simeq \frac{v_{in}}{\lambda_{qs}}  \exp -\frac{R_i
\sqrt{S/\mu^*_{min}}}{\lambda_{qs}}\left(\sqrt{\mu^*_{min}/\mu^*}-1\right).
\end{equation}

\noindent from which we obtain:

\begin{equation}\label{eqn:muIqs}
\mu^*(I) \simeq \mu^*_{min} \left(1 - \frac{\lambda_{qs}} {R_i
\sqrt{S/\mu^*_{min}}}\ln\left(\frac{\lambda_{qs}}{v_{in}}I\right)\right)^{-2}.
\end{equation}

\begin{figure}[!htb]
\begin{center}
\includegraphics*[width=7cm]{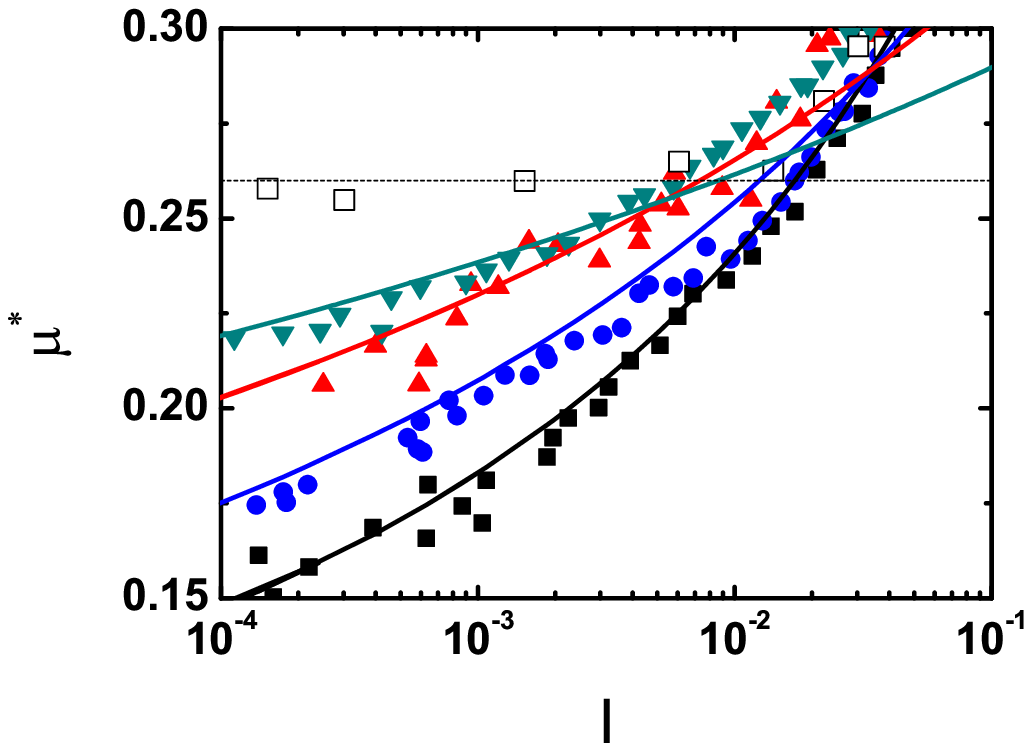}
\caption{\label{Fig25} \textit{Effective friction $\mu^*$ as
function of the inertial number $I$, in the quasistatic zone.
Comparison between the measurements: ($\blacksquare$) $R_{25}$,
($\textcolor[rgb]{0.00,0.00,1.00}{\bullet}$) $R_{50}$,
($\textcolor[rgb]{0.98,0.00,0.00}{\blacktriangle}$) $R_{100}$,
($\textcolor[rgb]{0.25,0.50,0.50}{\blacktriangledown}$) $R_{200}$,
($\square$) plane shear~\cite{Dacruz05} and the prediction of
Eqn.~\eqref{eqn:muIqs} (solid lines). The dashed line indicates
$\mu^*_{min}=0.26$. $V_{\theta}=2.5$.}}
\end{center}
\end{figure}

\noindent This prediction is in close agreement with the
measurements, as shown in Fig.~\ref{Fig25}.

\section{Conclusion}

We first summarize the results presented in this paper, before
discussing the questions raised by those conclusions.

As described in Sec.~\ref{sec:system}, we have studied through
discrete simulations steady annular shear flows of a model
granular material, made of a slightly polydisperse assembly of
frictional dissipative disks, prescribing the rotation rate of the
inner wall and the pressure exerted by the outer wall, and varying
dimensionless shear velocity $V_{\theta}$ and size $R_i$ of the
system.

The first step (Sec.~\ref{sec:localization}) has consisted in
measuring various quantities, either global as the dimensionless
shear stress at the inner wall $S$ as a function of $V_{\theta}$,
or local as the profiles of stress components, velocity, solid
fraction and some internal variables (coordination number,
mobilization of friction, velocity fluctuations, shown in App.~C).
This has allowed to distinguish, at the global scale, that is to
say as a function of $V_{\theta}$, between rate dependent and rate
independent behaviors.

Inspired by our previous rheophysical analysis of homogenous shear
flows of disks~\cite{Dacruz05}, the second step
(Sec.~\ref{sec:behavior}) has explored the validity of
constitutive law for inertial regime if applied locally in such an
heterogeneously sheared material. We have shown that the dynamic
friction and dilatancy laws observed in homogeneous shear flows
are exactly recovered, when using the local state parameter $I$
called \emph{inertial number}. Scaling laws for internal variables
as function of $I$ have also been observed. This analysis has
clearly distinguished an inertial zone close to the inner wall
where the constitutive law is relevant and a quasistatic zone away
from it, where it fails.

The last step (Sec.~\ref{sec:prediction}) has explained how it is
possible to predict some observations presented in the first step,
when using the inertial constitutive law identified in the second
step. We have focused on two basic quantities, which are most
often discussed in the studies of annular shear flows of granular
materials, the macroscopic $S(V_{\theta})$ relation and the
microscopic velocity profiles. The satisfactory agreement between
the prediction and the measurements should not be surprising,
considering the second step. However this analysis has precisely
pointed out two important issues, related to boundary conditions
in such a heterogeneously sheared system, one at the shearing
wall, and the other at the transition between inertial and
quasistatic zone.

Close to the inner wall, as previously shown in different
configurations, various quantities (solid fraction, ratio of
normal stresses, rotation velocity...) present singular behaviors.
The translation velocity reveals significant sliding for
sufficiently high $V_{\theta}$, even with the large roughness used
in this study. We have shown that the value $V_{\theta}^+$ of this
sliding velocity is an important ingredient for a good prediction.
This means that a detailed understanding of the rheophysics of the
granular materials in the very first layers near a rough wall is
of great relevance. Apart from the characteristics of the granular
material itself, the relative influences of $V_{\theta}$, $R_i$
and the wall roughness must be taken into account. Comparisons
between physical experiments and discrete simulations are
described in~\cite{Koval08a}. Considering the frustration of the
particle rotation imposed by the wall, Cosserat models might be
adapted to describe this interface zone~\cite{Tejchman93b}, as
done by~\cite{Latzel00,Mohan02} for annular shear. Another
discussion of the boundary condition at the wall is proposed
in~\cite{Artoni08}.

The transition between inertial and quasistatic zones is a second
puzzling issue. Considering the constitutive law identified in
homogeneous shear flows, the granular material should reach the
so-called critical state in the quasistatic limit (when $I \to
0$), in which it flows rate independently with an effective
friction $\tan \phi$ and a solid fraction $\nu_c$. Beyond this
limit (for $S/P < \tan \phi$ and/or $\nu > \nu_c$), the granular
material, being in a solid-like state, should not be able to flow.
However (apparently unbounded) creep flows are observed in this
nominally solid regime. This creeping behavior is well known in
free surface flows, where an exponential velocity profile has been
clearly evidenced with a characteristic length of the order of one
grain
diameter~\cite{Rajchenbach00,Khakhar01,Komatsu01,Crassous08}. In
the annular shear geometry, a similar behavior is observed but the
characteristic length increases as $R_i$ increases, that is to say
as the stress gradient decreases, or as the stress field becomes
more homogeneous.

For a sufficiently small $V_{\theta}$, there is no more inertial
zone, so that both boundary conditions occur at the same place,
the inner wall. Considering the typical values of the parameters
in the systems which have been studied experimentally or through
discrete simulations, we notice that this corresponds to the usual
case. The understanding of such a situation merges the two
previous problems: the behavior of a granular material close to an
interface and in the quasistatic regime, together with the
heterogeneity of the stress field.

The already noticed observation of collective and intermittent
motions in this quasistatic regime has driven the development of
several rheological models
(see~\cite{Debregeas00,Bocquet02b,Tardos03,Lagree06}
and~\cite{Mills08} for a recent review): diffusion equation for
the fluctuations, transmission of forces at the scale of
correlated clusters, two-phase fluid model with order parameter,
activation of rearrangements through the fluctuations of velocity
or forces, occurring either at the boundary of the inertial zone,
or at the inner wall in the global quasistatic limit.

Our understanding is far from complete and requires further
studies, merging physical experiments, discrete simulations and
theoretical developments. For instance, we have not measured the
fabric in the quasistatic zone, although its importance has been
clearly evidenced in homogeneous shear~\cite{Radjai04, Radjai08}.
We have not discussed the evolution of the internal variables in
the transient regime (evolution from initial to steady state), or
in a shear reversal regime~\cite{Losert01,Utter04a}, as should be
qualitatively possible using simplified microscopic
description~\cite{Falk08}. We have restricted our attention to
velocity controlled shear flows, so that it was not possible to
study the flow threshold. A specific study of the jamming
mechanisms should be performed by controlling the shear
stress~\cite{Dacruz02}. We have not discussed the influence of the
roughness on the interface behavior, for which we refer
to~\cite{Koval08a}. We may also wonder to what extent the
conclusions drawn for granular materials differ for other complex
fluids made of interacting elements (dense suspensions, foam,
emulsions...)~\cite{Debregeas01,Huang05}.

\section*{Appendix A : Periodic boundary condition} \label{app:a}

Each grain the center of which is in $(r,\theta)$ with $0 \leq
\theta \leq \Theta$ is associated to a collection of copies with
centers in $r,\theta+k\Theta$ where $k$ is an integer. The
corresponding velocities, accelerations and forces are related by
rotations of angles $k\Theta$.

Every time a grain moves out of the simulation cell, one of its
copies moves in by the opposite boundary, similarly to the usual
case of periodic boundary conditions by translation. However the
velocities, accelerations and forces are affected by a rotation of
$\pm \Theta$.

\begin{figure}[!htb]
\begin{center}
\includegraphics*[width=8cm]{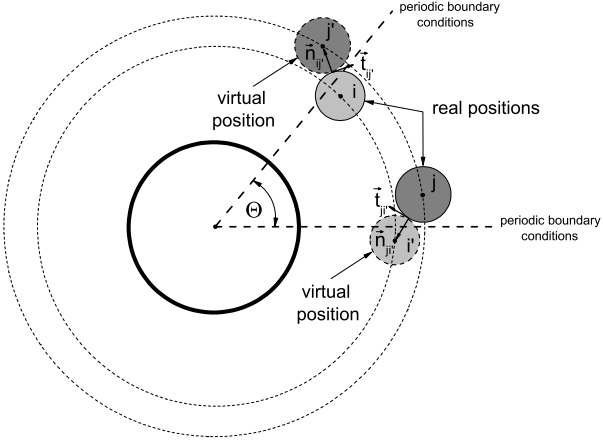}
\caption{\label{Fig26} \textit{Periodic boundary conditions.}}
\end{center}
\end{figure}

The situation of the contact of two grains $i$ and $j$ where
$\theta_i$ is close to $\Theta$ and $\theta_j$ is close to zero
is described in Fig.~\ref{Fig26}. More precisely, $i$ is in
contact with the copy $j'$ of $j$, obtained by rotation of an
angle $\Theta$, while $j$ is in contact with $i'$ obtained by
rotation of $i$ of an angle $-\Theta$. To evaluate the forces
acting over grain $i$ we have to use the normal and tangential
unit vectors $\vec{n}_{ij'}$ (pointing from $i$ to $j'$) and
$\vec{t}_{ij'}$ (such that $({\vec n}_{ij'},{\vec t}_{ij'})$ is
positively oriented), respectively, and the motion of the grain
$j'$, while for $j$ we have to use corresponding $\vec{n}_{ji'}$
and $\vec{t}_{ji'}$ and the motion of $i'$. Vector
$\vec{n}_{ij'}$ is not, as usually, equal to $-\vec{n}_{ji'}$,
but to its image obtained by rotation of an angle $-\Theta$.

\begin{figure}[!htb]
\begin{center}
\includegraphics*[width=7cm]{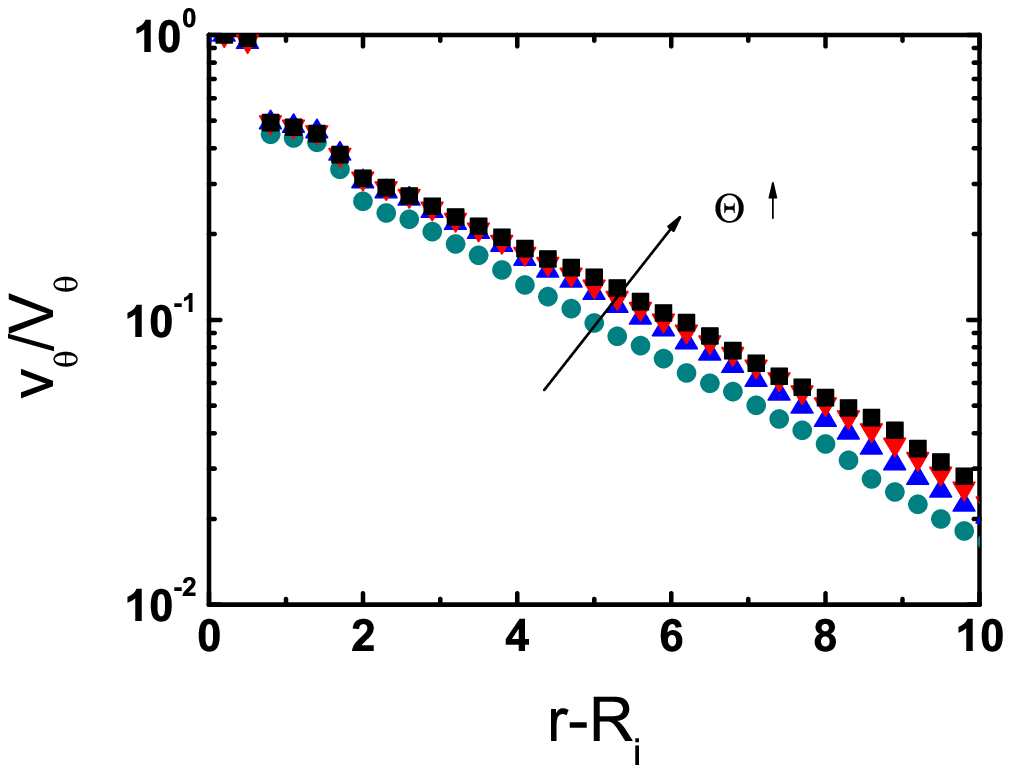}
\caption{\label{Fig27} \textit{(Color online) Velocity profiles
$v_{\theta}(r)/V_{\theta}$ for different values of $\Theta$
(rad). ($\textcolor[rgb]{0.25,0.50,0.50}{\bullet}$)
$\Theta=\pi/16$,
($\textcolor[rgb]{0.00,0.00,1.00}{\blacktriangle}$)
$\Theta=\pi/8$,
($\textcolor[rgb]{0.98,0.00,0.00}{\blacktriangledown}$)
$\Theta=\pi/2$, ($\blacksquare$) $\Theta=2 \pi$. $R_{i}=25$,
$R_{o}=50$, $V_{\theta}=2.5$.}}
\end{center}
\end{figure}

We have measured the influence of the periodic boundary condition
comparing the radial profiles of various quantities as a function
of $\Theta$ ($\pi /16$, $\pi /8$, $\pi /4$, $\pi /2$, $\pi$ and
the whole ring $2 \pi$) for the geometry $R_i=25$ and $R_o=50$.
As an example, we show on Fig.~\ref{Fig27} the profiles of the
orthoradial velocity. As expected, the results are all the more
consistent as the value of $\Theta$ increases. In this case,
$\Theta=\pi/2$ already gives a very good result.

We quantify the deviations of the velocity profiles
$v_{\theta}(r)$ by means of an indicator of relative error. The
velocity tends to zero as the distance from the inner wall. To
avoid inconsistencies due to values close to zero in the frame of
the usual definition of relative error, and to give more weight to
the values close to the inner wall, we propose to calculate the
relative error over variable
$F_{\Theta}(r)=V_{\theta}-v_{\theta}(r,\Theta)$ :

\begin{equation}\label{eqn:indic}
\ \varepsilon(\Theta)= \frac{1}{R_{o}-R_{i}}\int^{R_{o}}_{R_{i}}|{
\frac{F_{\Theta}(r)-F_{2\Theta}(r)}{F_{2\Theta}(r)}}|dr.
\end{equation}

\noindent $\varepsilon(\Theta)$ is simply the sum over the whole
geometry of the relative error of the variable $F$ for a certain
value of $\Theta$ compared to the result for a system twice as
large ($2\Theta$).

\begin{figure}[!htb]
\begin{center}
\includegraphics*[width=7cm]{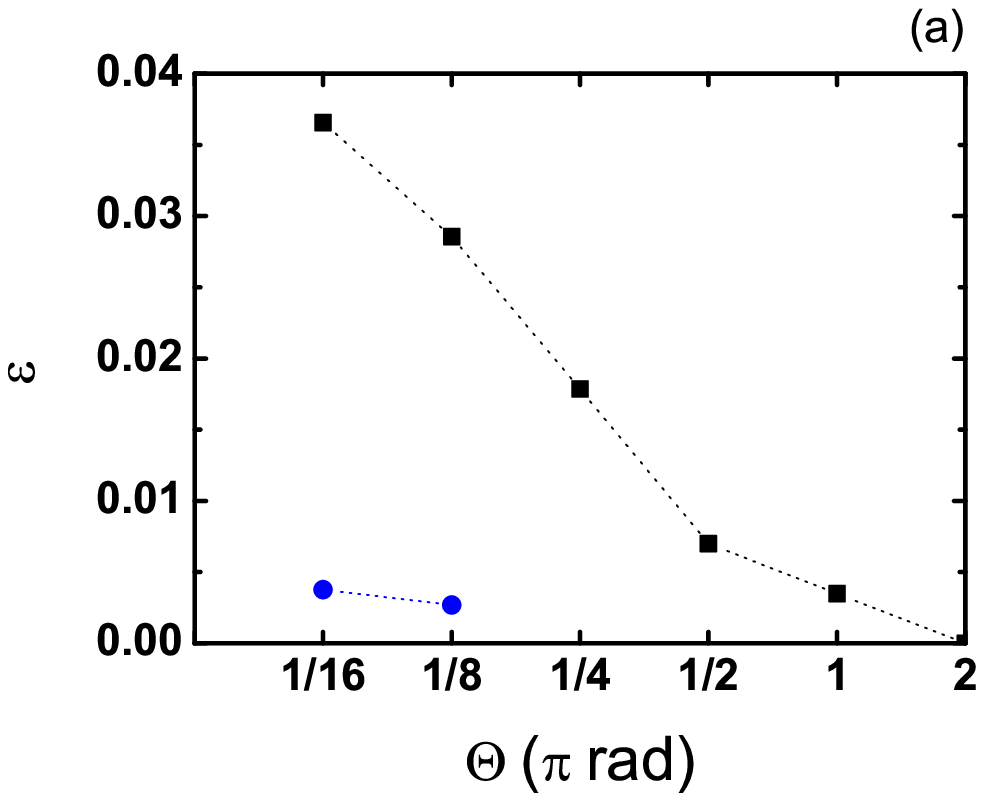}
\includegraphics*[width=7cm]{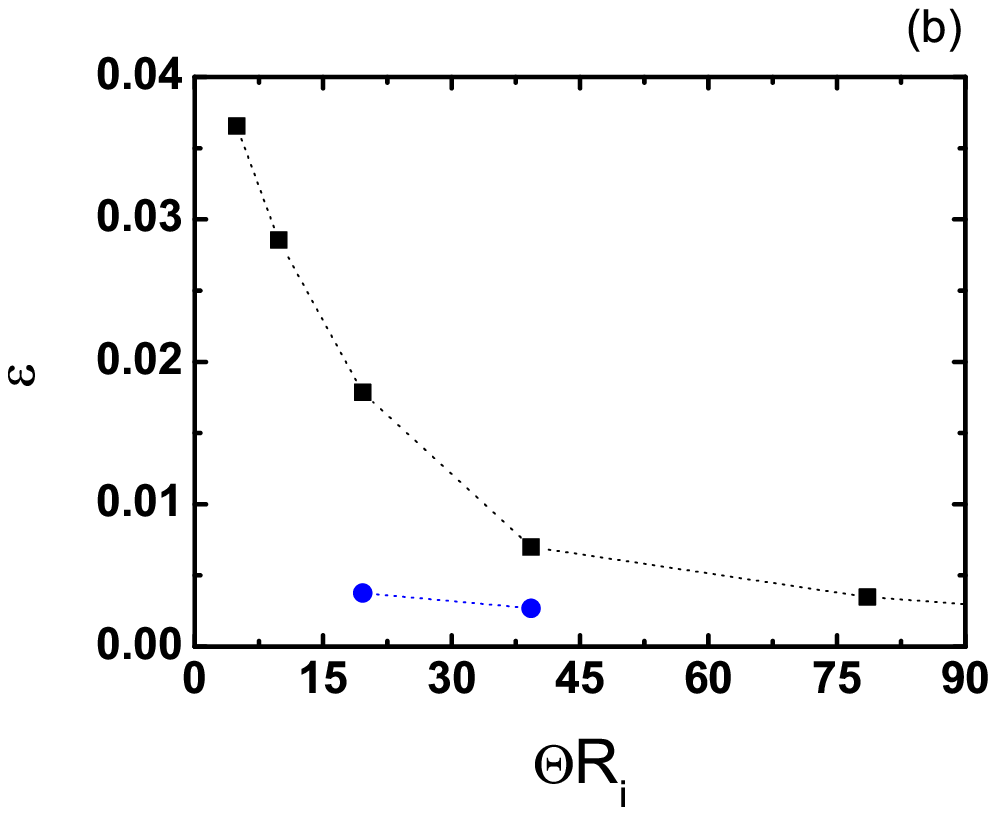}
\caption{\label{Fig28} \textit{(Color online) Relative error
$\varepsilon$ on the orthoradial velocity $v_{\theta}$
($V_{\theta}=$~2.5), (a) as function of $\Theta$, (b) as function
of the inner wall length $\Theta R_{i}$. ($\blacksquare$)
$R_{i}=25$ and $R_{o}=50$,
($\textcolor[rgb]{0.00,0.00,1.00}{\bullet}$) $R_{i}=100$ and
$R_{o}=200$.}}
\end{center}
\end{figure}

On Fig.~\ref{Fig28}a, we observe a clear decrease of the error
indicator $\varepsilon(\Theta)$ as we increase the value of
$\Theta$ for the smallest geometry ($R_i=25$ and $R_o=50$). The
same analysis for a larger geometry ($R_i=100$ and $R_o=200$)
shows better results for smaller values of $\Theta$. This shows
that the influence of $\Theta$ on the results depends on the size
of the system. We try to relate both parameters in
Fig.~\ref{Fig28}b, where we plot $\varepsilon(\Theta)$ as a
function of the angular sector length at the inner wall ($\Theta
R_i$). We observe that a good accuracy of the results can be
achieved with a length $\Theta R_i \geq 40$ for geometries with
$R_i \geq 25$. Based on this consideration, we have chosen the
values of $\Theta$ for each of our geometries (Tab.~I).

\section*{Appendix B : Averaging method} \label{app:b}

\begin{figure}[!htb]
\begin{center}
\includegraphics*[width=7cm]{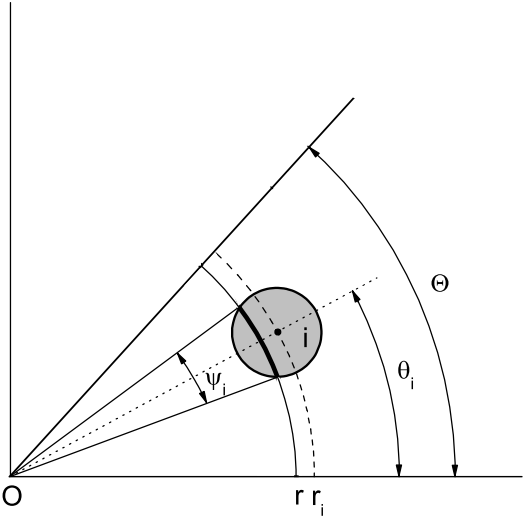}
\caption{\label{Fig29} \textit{Various quantities associated to a
grain $i$.}}
\end{center}
\end{figure}

Considering the revolution symmetry of our system, the radial
profiles of different quantities (orthoradial velocity
$v_{\theta}(r)$, coordination number $Z(r)$, etc.) are obtained
by an averaging procedure over coordinate $\theta$ along the
coordinate $r$ (Fig.~\ref{Fig29}). To each of the $n$ grains $i$
are associated different scalar quantities $G^i$. We define a
weight function $\psi_i(r)$ as the intercept angle defined on
Fig.~\ref{Fig29} ($\cos (\psi_i(r)/2) =
(r^2+r_i^2-d_i^2/4)/(2rr_i)$ for a disk of diameter $d_i$). Some
variables, like solid fraction $\nu$, are averaged over the whole
space, while others, like the coordination number $Z$, have no
sense outside the grain space. This leads to the two following
definitions of the average :

\begin{equation}
    <G>(r) = \frac{1}{\Theta} \sum_{i=1}^n G^i \psi_i(r),
\end{equation}

\noindent and :

\begin{equation}
    <G>'(r) = \frac{\sum_{i=1}^n G^i \psi_i(r)}{\sum_{i=1}^n
    \psi_i(r)}.
\end{equation}

\noindent

Applying this principle, we determine the solid fraction profile
$\nu(r)$ as follows

\begin{equation}
    <\nu>(r) = \frac{1}{\Theta} \sum_{i=1}^n  \psi_i(r),
\end{equation}

\noindent where the value of $\nu_i$ is naturally equal to $1$.
This means that $<G>$ and $<G>'$ are simply related by the solid
fraction: $<G>=<\nu><G>'$.

We take into account the variation of vectorial and tensorial
quantities inside the grains, when written in the polar basis $\vec{e}_r(\phi)=\left(%
\begin{array}{c}
  \cos \phi \\
  \sin \phi \\
\end{array}%
\right)$ and $\vec{e}_{\theta}=\left(%
\begin{array}{c}
  -\sin \phi \\
  \cos \phi \\
\end{array}%
\right)$. Hence, the radial profiles of the velocity components
are :

\begin{equation}
v_{\alpha}(r) = \frac{1}{\sum_{i=1}^n \psi_i(r)} \sum_{i=1}^n
\int^{\theta_i+\frac{\psi_i}{2}}_{\theta_i-\frac{\psi_i}{2}}
\vec{v}^i \cdot \vec{e}_{\alpha}(\phi) d\phi.
\end{equation}

The stress tensor of each grain is defined according
to~\cite{Moreau97} (with $A^i=\pi d_i^2/4$ the grain area) :

\begin{equation}
\underline{\underline{\sigma}}^i = \frac{1}{A^i} \left(
    \sum_{j \ne i} {\vec F}_{ij} \otimes {\vec r}_{ij} + m_{i} \delta \vec v_{i} \otimes \delta \vec
    v_{i} \right).
\end{equation}

\noindent The first term is associated to the contact forces, and
the second one to the velocity fluctuations. The radial profiles
of the components of the stress tensor are :

\begin{equation}
\sigma_{\alpha \beta}(r) = \frac{1}{\Theta} \sum_{i=1}^n
\int^{\theta_i+\frac{\psi_i}{2}}_{\theta_i-\frac{\psi_i}{2}}
\vec{e}_{\alpha}(\phi) \cdot \underline{\underline{\sigma}}^i
\cdot \vec{e}_{\beta}(\phi)d\phi.
\end{equation}

Since we try to analyze the granular material as a continuum
(except for the very first layers near the wall), we consider the
coarse-grained variations of the quantities by smoothing the
profiles through central moving averages of $3d$ length (if not
otherwise indicated). The remaining fluctuations would disappear
with an increase of the simulation time $\Delta t$ over which the
data are averaged.\\

\section*{Appendix C : Internal variables} \label{app:c}

\begin{figure}[!htb]
\begin{center}
\includegraphics*[width=7cm]{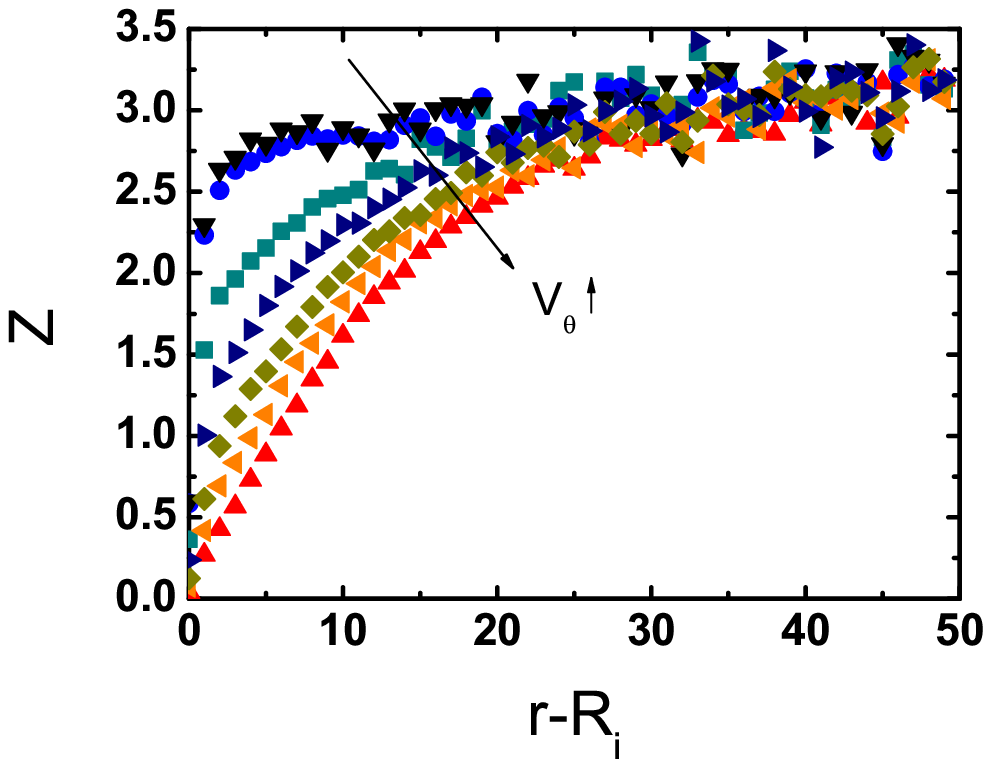}
\caption{\label{Fig8} \textit{(Color online) Influence of shear
velocity $V_{\theta}$ on the coordination number profiles $Z(r)$.
($\blacktriangledown$) $V_{\theta}=0.0025$,
($\textcolor[rgb]{0.00,0.00,1.00}{\bullet}$) $V_{\theta}=0.025$,
($\textcolor[rgb]{0.25,0.50,0.50}{\blacksquare}$)
$V_{\theta}=0.25$,
($\textcolor[rgb]{0.00,0.00,0.50}{\blacktriangleright}$)
$V_{\theta}=0.5$,
($\textcolor[rgb]{0.50,0.50,0.00}{\blacklozenge}$)
$V_{\theta}=1.0$,
($\textcolor[rgb]{1.00,0.50,0.00}{\blacktriangleleft}$)
$V_{\theta}=1.5$,
($\textcolor[rgb]{0.98,0.00,0.00}{\blacktriangle}$)
$V_{\theta}=2.5$. Geometry $R_{50}$.}}
\end{center}
\end{figure}

Coordination number $Z$ is the average number of contacts per
grain. In the inertial regime, the general tendency is a decrease
of $Z$ as the shear rate $\dot{\gamma}$ increases (that is to say
for increasing $V_{\theta}$ in Fig.~\ref{Fig8}). For smaller
values of $\dot{\gamma}$ (corresponding to smaller values of
$V_{\theta}$ or to a larger distance from the inner wall) the
coordination number $Z$ approaches a limiting value, slightly
above 3. Such a limit is in rough agreement with other numerical
observations of the critical state of frictional disks.
Ref.~\cite{Radjai04} thus reports $Z \simeq 3.6$. The somewhat
lower values observed in our case are likely due to the larger
strain rates, and to the remaining influence, on the quasistatic
region of limited width, of the more agitated inner zone.

\begin{figure}[!htb]
\begin{center}
\includegraphics*[width=7cm]{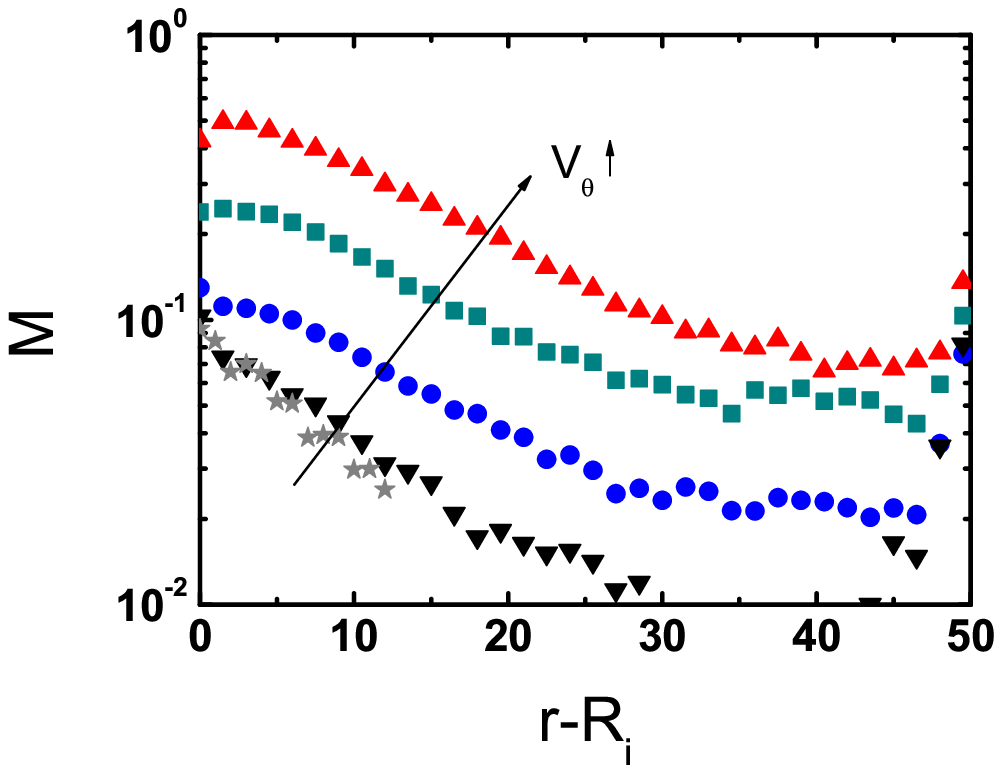}
\caption{\label{Fig10} \textit{(Color online) Influence of shear
velocity $V_{\theta}$ on the mobilization of friction profiles
$M(r)$. ($\textcolor[rgb]{0.50,0.50,0.50}{\bigstar}$)
$V_{\theta}=0,00025$, ($\blacktriangledown$) $V_{\theta}=0.0025$,
($\textcolor[rgb]{0.00,0.00,1.00}{\bullet}$) $V_{\theta}=0.025$,
($\textcolor[rgb]{0.25,0.50,0.50}{\blacksquare}$)
$V_{\theta}=0.25$,
($\textcolor[rgb]{0.98,0.00,0.00}{\blacktriangle}$)
$V_{\theta}=2.5$. Geometry $R_{50}$.}}
\end{center}
\end{figure}

We define the mobilization of friction as ratio $M=Z_{s}/Z$, where
$Z_{s}$ is the average number of \emph{sliding} contacts per
grain~\cite{Staron02b,Dacruz05}. Fig.~\ref{Fig10} shows that $M$
increases as the shear rate increases, whether through an increase
of $V_{\theta}$ or a decrease of the distance from the inner wall.
We notice that the stabilization of the $M(r)$ profile occurs for
$V_{\theta}\leq 0.0025$, a value much smaller than the one
required for the stabilization of the other studied quantities
($V_{\theta}\leq 0.025$).

For any quantity $q(r)$ averaged in space (along $\theta$) and in
time, we may define its fluctuation:

\begin{eqnarray}
\delta q(r)^2 & = & \frac{1}{\Theta}\int_0^{\Theta}
q(r,\theta)^{2}d\theta - q(r)^{2},
\end{eqnarray}

\noindent where $q(r,\theta)^{2}$ is averaged in time. We measure
the fluctuations of the translational and rotational velocities
$\delta v_{\theta}(r)$, $\delta v_{r}(r)$ and $\delta \omega(r)$.
Our analysis (long time scale) takes into account both the small
fluctuations around the mean motion (in the \emph{cage} formed by
the nearest neighbors), and the large fluctuations associated to
collective motions~\cite{Radjai02}.

\begin{figure}[!htb]
\begin{center}
\includegraphics*[width=7cm]{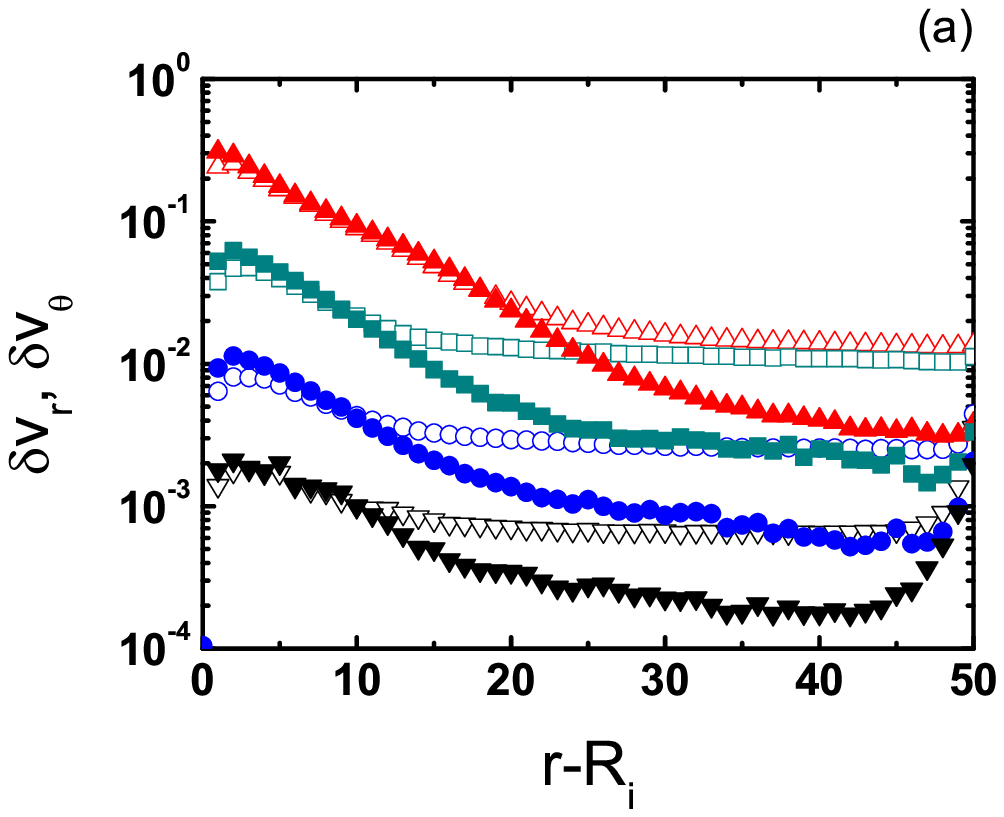}
\includegraphics*[width=7cm]{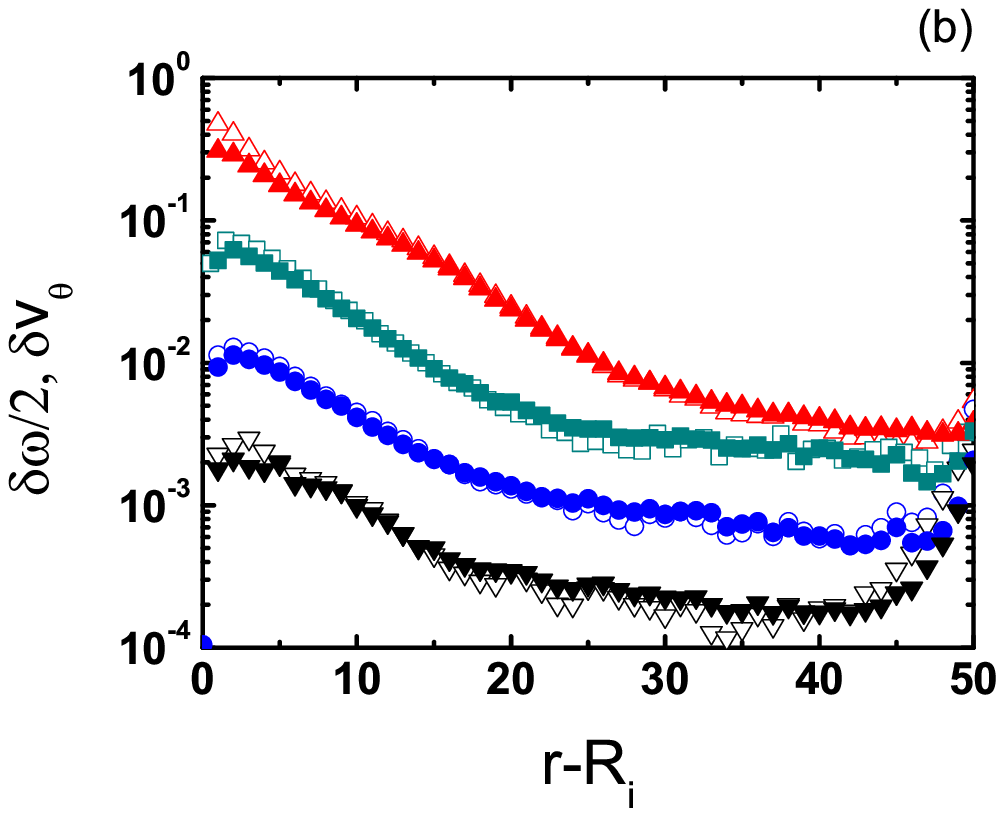}
\caption{\label{Fig11} \textit{(Color online) (a) Comparison
between the profiles of the fluctuations of the radial velocity
$\delta v_r$ (hollow symbols) and of the orthoradial velocity
$\delta v_{\theta}$ (full symbols). (b) Comparison between the
profiles of the fluctuations of the angular velocity $\delta
\omega/2$ (hollow symbols) and of the orthoradial velocity $\delta
v_{\theta}$ (full symbols) for different wall velocities
$V_{\theta}$: ($\triangledown$, $\blacktriangledown$)
$V_{\theta}=0.0025$, ($\textcolor[rgb]{0.00,0.00,1.00}{\circ ,
\bullet}$) $V_{\theta}=0.025$,
($\textcolor[rgb]{0.25,0.50,0.50}{\square, \blacksquare}$)
$V_{\theta}=0.25$, ($\textcolor[rgb]{0.98,0.00,0.00}{\vartriangle,
\blacktriangle}$) $V_{\theta}=2.5$. Geometry $R_{50}$.}}
\end{center}
\end{figure}

Fig.~\ref{Fig11} first shows that the general amplitude of the
fluctuations increases with $V_{\theta}$. Then, for various
$V_{\theta}$, they reveal a strong decay of the fluctuating
quantities close to the inner wall, comparable to that of the
respective average quantities, consistently with previous
observations~\cite{Denniston99,Schollmann99,Veje99,Mueth00,Mueth03,Dacruz05}.
This decay is still true at larger distances for $\delta
v_{\theta}$ and $\delta \omega$ (with an increase close to the
outer wall). We also notice a stabilization of $\delta v_{r}$,
which occurs at $r-R_i \approx 10$ for $V_{\theta}=0.025$ and at
$r-R_i \approx 20$ for $V_{\theta}=2.5$, that is to say precisely
when the solid fraction $\nu$ reaches a value $\approx 0.82$
(Fig.~\ref{Fig7}). Above this critical value of $\nu$, the
material would be so compact that the radial motions would take
place as a block. Fig.~\ref{Fig11}a shows the equality of $\delta
v_{\theta}$ and $\delta v_{r}$ before the stabilization of $\delta
v_{r}(r)$, while Fig.~\ref{Fig11}b shows the systematic equality
of $\delta v_{\theta}$ and $\delta \omega/2$.

\section*{Acknowledgments}

We gratefully acknowledge Frédéric da Cruz, Anaël Lemaître, Jean
Sulem and Dietrich Wolf for many interesting discussions at
various stages of this study.

Institut Navier is a joint laboratory, depending on Laboratoire
Central des Ponts et Chaussées, Ecole Nationale des Ponts et
Chaussées and Centre National de la Recherche Scientifique.


\begin{thebibliography}{76}
\expandafter\ifx\csname
natexlab\endcsname\relax\def\natexlab#1{#1}\fi
\expandafter\ifx\csname bibnamefont\endcsname\relax
  \def\bibnamefont#1{#1}\fi
\expandafter\ifx\csname bibfnamefont\endcsname\relax
  \def\bibfnamefont#1{#1}\fi
\expandafter\ifx\csname citenamefont\endcsname\relax
  \def\citenamefont#1{#1}\fi
\expandafter\ifx\csname url\endcsname\relax
  \def\url#1{\texttt{#1}}\fi
\expandafter\ifx\csname
urlprefix\endcsname\relax\def\urlprefix{URL }\fi
\providecommand{\bibinfo}[2]{#2}
\providecommand{\eprint}[2][]{\url{#2}}

\bibitem[{\citenamefont{Pouliquen and Chevoir}(2002)}]{Pouliquen02a}
\bibinfo{author}{\bibfnamefont{O.}~\bibnamefont{Pouliquen}} \bibnamefont{and}
  \bibinfo{author}{\bibfnamefont{F.}~\bibnamefont{Chevoir}},
  \bibinfo{journal}{Comptes-Rendus Physique} \textbf{\bibinfo{volume}{3}},
  \bibinfo{pages}{163} (\bibinfo{year}{2002}).

\bibitem[{\citenamefont{{GDR MIDI}}(2004)}]{Gdr04}
\bibinfo{author}{\bibnamefont{{GDR MIDI}}}, \bibinfo{journal}{Euro. Phys. J. E}
  \textbf{\bibinfo{volume}{14}}, \bibinfo{pages}{341} (\bibinfo{year}{2004}).

\bibitem[{\citenamefont{Forterre and Pouliquen}(2008)}]{Forterre08}
\bibinfo{author}{\bibfnamefont{Y.}~\bibnamefont{Forterre}} \bibnamefont{and}
  \bibinfo{author}{\bibfnamefont{O.}~\bibnamefont{Pouliquen}},
  \bibinfo{journal}{Annu. Rev. Fluid Mech.} \textbf{\bibinfo{volume}{40}},
  \bibinfo{pages}{1} (\bibinfo{year}{2008}).

\bibitem[{\citenamefont{da~Cruz et~al.}(2005)\citenamefont{da~Cruz, Emam,
  Prochnow, Roux, and Chevoir}}]{Dacruz05}
\bibinfo{author}{\bibfnamefont{F.}~\bibnamefont{da~Cruz}},
  \bibinfo{author}{\bibfnamefont{S.}~\bibnamefont{Emam}},
  \bibinfo{author}{\bibfnamefont{M.}~\bibnamefont{Prochnow}},
  \bibinfo{author}{\bibfnamefont{J.-N.} \bibnamefont{Roux}}, \bibnamefont{and}
  \bibinfo{author}{\bibfnamefont{F.}~\bibnamefont{Chevoir}},
  \bibinfo{journal}{Phys. Rev. E} \textbf{\bibinfo{volume}{72}},
  \bibinfo{pages}{021309} (\bibinfo{year}{2005}).

\bibitem[{\citenamefont{Lois et~al.}(2005)\citenamefont{Lois, Lemaître, and
  Carlson}}]{Lois05}
\bibinfo{author}{\bibfnamefont{G.}~\bibnamefont{Lois}},
  \bibinfo{author}{\bibfnamefont{A.}~\bibnamefont{Lemaître}}, \bibnamefont{and}
  \bibinfo{author}{\bibfnamefont{J.~M.} \bibnamefont{Carlson}},
  \bibinfo{journal}{Phys. Rev. E} \textbf{\bibinfo{volume}{72}},
  \bibinfo{pages}{051303} (\bibinfo{year}{2005}).

\bibitem[{\citenamefont{Chevoir et~al.}(2008)\citenamefont{Chevoir, Roux,
  da~Cruz, Rognon, and Koval}}]{Chevoir08b}
\bibinfo{author}{\bibfnamefont{F.}~\bibnamefont{Chevoir}},
  \bibinfo{author}{\bibfnamefont{J.-N.} \bibnamefont{Roux}},
  \bibinfo{author}{\bibfnamefont{F.}~\bibnamefont{da~Cruz}},
  \bibinfo{author}{\bibfnamefont{P.}~\bibnamefont{Rognon}}, \bibnamefont{and}
  \bibinfo{author}{\bibfnamefont{G.}~\bibnamefont{Koval}},
  \bibinfo{journal}{Powder Tech.}  (\bibinfo{year}{2008}),
  \bibinfo{note}{doi:10.1016/j.powtec.2008.04.061}.

\bibitem[{\citenamefont{Jop et~al.}(2006)\citenamefont{Jop, Forterre, and
  Pouliquen}}]{Jop06}
\bibinfo{author}{\bibfnamefont{P.}~\bibnamefont{Jop}},
  \bibinfo{author}{\bibfnamefont{Y.}~\bibnamefont{Forterre}}, \bibnamefont{and}
  \bibinfo{author}{\bibfnamefont{O.}~\bibnamefont{Pouliquen}},
  \bibinfo{journal}{Nature} \textbf{\bibinfo{volume}{441}},
  \bibinfo{pages}{727} (\bibinfo{year}{2006}).

\bibitem[{\citenamefont{Hatano}(2007)}]{Hatano07}
\bibinfo{author}{\bibfnamefont{T.}~\bibnamefont{Hatano}},
  \bibinfo{journal}{Phys. Rev. E} \textbf{\bibinfo{volume}{75}},
  \bibinfo{pages}{060301} (\bibinfo{year}{2007}).

\bibitem[{\citenamefont{Peyneau and Roux}(2008)}]{Peyneau08}
\bibinfo{author}{\bibfnamefont{P.-E.} \bibnamefont{Peyneau}} \bibnamefont{and}
  \bibinfo{author}{\bibfnamefont{J.-N.} \bibnamefont{Roux}},
  \bibinfo{journal}{Phys. Rev. E} \textbf{\bibinfo{volume}{78}},
  \bibinfo{pages}{011307} (\bibinfo{year}{2008}).

\bibitem[{\citenamefont{Goldhirsch}(2003)}]{Goldhirsch03}
\bibinfo{author}{\bibfnamefont{I.}~\bibnamefont{Goldhirsch}},
  \bibinfo{journal}{Annu. Rev. Fluid Mech.} \textbf{\bibinfo{volume}{35}},
  \bibinfo{pages}{267} (\bibinfo{year}{2003}).

\bibitem[{\citenamefont{{Schofield} and {Wroth}}(1968)}]{Schofield68}
\bibinfo{author}{\bibfnamefont{A.~N.} \bibnamefont{{Schofield}}}
  \bibnamefont{and} \bibinfo{author}{\bibfnamefont{C.~P.}
  \bibnamefont{{Wroth}}}, \emph{\bibinfo{title}{Critical state soil mechanics}}
  (\bibinfo{publisher}{McGraw-Hill}, \bibinfo{address}{London},
  \bibinfo{year}{1968}).

\bibitem[{\citenamefont{Wood}(1990)}]{Wood90}
\bibinfo{author}{\bibfnamefont{D.~M.} \bibnamefont{Wood}},
  \emph{\bibinfo{title}{Soil Behaviour and Critical State Soil Mechanics}}
  (\bibinfo{publisher}{Cambridge University Press}, \bibinfo{year}{1990}).

\bibitem[{\citenamefont{de~Ryck et~al.}(2008)\citenamefont{de~Ryck, Ansart, and
  Dodds}}]{Deryck08a}
\bibinfo{author}{\bibfnamefont{A.}~\bibnamefont{de~Ryck}},
  \bibinfo{author}{\bibfnamefont{R.}~\bibnamefont{Ansart}}, \bibnamefont{and}
  \bibinfo{author}{\bibfnamefont{J.~A.} \bibnamefont{Dodds}},
  \bibinfo{journal}{Granular Matter} \textbf{\bibinfo{volume}{10}},
  \bibinfo{pages}{353} (\bibinfo{year}{2008}).

\bibitem[{\citenamefont{Jop et~al.}(2005)\citenamefont{Jop, Forterre, and
  Pouliquen}}]{Jop05b}
\bibinfo{author}{\bibfnamefont{P.}~\bibnamefont{Jop}},
  \bibinfo{author}{\bibfnamefont{Y.}~\bibnamefont{Forterre}}, \bibnamefont{and}
  \bibinfo{author}{\bibfnamefont{O.}~\bibnamefont{Pouliquen}},
  \bibinfo{journal}{J. Fluid Mech.} \textbf{\bibinfo{volume}{541}},
  \bibinfo{pages}{167} (\bibinfo{year}{2005}).

\bibitem[{\citenamefont{Miller et~al.}(1996)\citenamefont{Miller, O'Hern, and
  Behringer}}]{Miller96}
\bibinfo{author}{\bibfnamefont{B.}~\bibnamefont{Miller}},
  \bibinfo{author}{\bibfnamefont{C.}~\bibnamefont{O'Hern}}, \bibnamefont{and}
  \bibinfo{author}{\bibfnamefont{R.~P.} \bibnamefont{Behringer}},
  \bibinfo{journal}{Phys. Rev. Lett.} \textbf{\bibinfo{volume}{77}},
  \bibinfo{pages}{3110} (\bibinfo{year}{1996}).

\bibitem[{\citenamefont{Elliott et~al.}(1998)\citenamefont{Elliott, Ahmadi, and
  Kvasnak}}]{Elliott98}
\bibinfo{author}{\bibfnamefont{K.~E.} \bibnamefont{Elliott}},
  \bibinfo{author}{\bibfnamefont{G.}~\bibnamefont{Ahmadi}}, \bibnamefont{and}
  \bibinfo{author}{\bibfnamefont{W.}~\bibnamefont{Kvasnak}},
  \bibinfo{journal}{J. Non Newtonian Fluid Mech.}
  \textbf{\bibinfo{volume}{74}}, \bibinfo{pages}{89} (\bibinfo{year}{1998}).

\bibitem[{\citenamefont{Veje et~al.}(1999)\citenamefont{Veje, Howell, and
  Behringer}}]{Veje99}
\bibinfo{author}{\bibfnamefont{C.}~\bibnamefont{Veje}},
  \bibinfo{author}{\bibfnamefont{D.}~\bibnamefont{Howell}}, \bibnamefont{and}
  \bibinfo{author}{\bibfnamefont{R.}~\bibnamefont{Behringer}},
  \bibinfo{journal}{Phys. Rev. E} \textbf{\bibinfo{volume}{59}},
  \bibinfo{pages}{739} (\bibinfo{year}{1999}).

\bibitem[{\citenamefont{Howell et~al.}(1999{\natexlab{a}})\citenamefont{Howell,
  Behringer, and Veje}}]{Howell99a}
\bibinfo{author}{\bibfnamefont{D.}~\bibnamefont{Howell}},
  \bibinfo{author}{\bibfnamefont{R.}~\bibnamefont{Behringer}},
  \bibnamefont{and} \bibinfo{author}{\bibfnamefont{C.}~\bibnamefont{Veje}},
  \bibinfo{journal}{Phys. Rev. Lett.} \textbf{\bibinfo{volume}{82}},
  \bibinfo{pages}{5241} (\bibinfo{year}{1999}{\natexlab{a}}).

\bibitem[{\citenamefont{Hartley and Behringer}(2003)}]{Hartley03}
\bibinfo{author}{\bibfnamefont{R.}~\bibnamefont{Hartley}} \bibnamefont{and}
  \bibinfo{author}{\bibfnamefont{R.}~\bibnamefont{Behringer}},
  \bibinfo{journal}{Nature} \textbf{\bibinfo{volume}{421}}, \bibinfo{pages}{928
  } (\bibinfo{year}{2003}).

\bibitem[{\citenamefont{{Tardos} et~al.}(1998)\citenamefont{{Tardos}, {Khan},
  and {Schaeffer}}}]{Tardos98}
\bibinfo{author}{\bibfnamefont{G.}~\bibnamefont{{Tardos}}},
  \bibinfo{author}{\bibfnamefont{M.}~\bibnamefont{{Khan}}}, \bibnamefont{and}
  \bibinfo{author}{\bibfnamefont{D.}~\bibnamefont{{Schaeffer}}},
  \bibinfo{journal}{Phys. Fluids} \textbf{\bibinfo{volume}{10}},
  \bibinfo{pages}{335} (\bibinfo{year}{1998}).

\bibitem[{\citenamefont{Howell et~al.}(1999{\natexlab{b}})\citenamefont{Howell,
  Behringer, and Veje}}]{Howell99b}
\bibinfo{author}{\bibfnamefont{D.}~\bibnamefont{Howell}},
  \bibinfo{author}{\bibfnamefont{R.}~\bibnamefont{Behringer}},
  \bibnamefont{and} \bibinfo{author}{\bibfnamefont{C.}~\bibnamefont{Veje}},
  \bibinfo{journal}{Chaos} \textbf{\bibinfo{volume}{9}}, \bibinfo{pages}{559}
  (\bibinfo{year}{1999}{\natexlab{b}}).

\bibitem[{\citenamefont{Mueth et~al.}(2000)\citenamefont{Mueth, Debregeas,
  Karczmar, Eng, Nagel, and Jaeger}}]{Mueth00}
\bibinfo{author}{\bibfnamefont{D.}~\bibnamefont{Mueth}},
  \bibinfo{author}{\bibfnamefont{G.}~\bibnamefont{Debregeas}},
  \bibinfo{author}{\bibfnamefont{G.}~\bibnamefont{Karczmar}},
  \bibinfo{author}{\bibfnamefont{P.}~\bibnamefont{Eng}},
  \bibinfo{author}{\bibfnamefont{S.}~\bibnamefont{Nagel}}, \bibnamefont{and}
  \bibinfo{author}{\bibfnamefont{H.}~\bibnamefont{Jaeger}},
  \bibinfo{journal}{Nature} \textbf{\bibinfo{volume}{406}},
  \bibinfo{pages}{385} (\bibinfo{year}{2000}).

\bibitem[{\citenamefont{{Klausner} et~al.}(2000)\citenamefont{{Klausner}, Chen,
  and Mei}}]{Klausner00}
\bibinfo{author}{\bibfnamefont{J.}~\bibnamefont{{Klausner}}},
  \bibinfo{author}{\bibfnamefont{D.}~\bibnamefont{Chen}}, \bibnamefont{and}
  \bibinfo{author}{\bibfnamefont{R.}~\bibnamefont{Mei}},
  \bibinfo{journal}{Powder Tech.} \textbf{\bibinfo{volume}{112}},
  \bibinfo{pages}{94} (\bibinfo{year}{2000}).

\bibitem[{\citenamefont{Cain}(2001)}]{Cain01}
\bibinfo{author}{\bibfnamefont{R.}~\bibnamefont{Cain}}, \bibinfo{journal}{Phys.
  Rev. E} \textbf{\bibinfo{volume}{64}}, \bibinfo{pages}{016413}
  (\bibinfo{year}{2001}).

\bibitem[{\citenamefont{Losert and Kwon}(2001)}]{Losert01}
\bibinfo{author}{\bibfnamefont{W.}~\bibnamefont{Losert}} \bibnamefont{and}
  \bibinfo{author}{\bibfnamefont{G.}~\bibnamefont{Kwon}},
  \bibinfo{journal}{Advances in Complex Systems} \textbf{\bibinfo{volume}{4}},
  \bibinfo{pages}{369} (\bibinfo{year}{2001}).

\bibitem[{\citenamefont{{Bocquet} et~al.}(2002)\citenamefont{{Bocquet},
  {Losert}, {Schalk}, {Lubensky}, and {Gollub}}}]{Bocquet02b}
\bibinfo{author}{\bibfnamefont{L.}~\bibnamefont{{Bocquet}}},
  \bibinfo{author}{\bibfnamefont{W.}~\bibnamefont{{Losert}}},
  \bibinfo{author}{\bibfnamefont{D.}~\bibnamefont{{Schalk}}},
  \bibinfo{author}{\bibfnamefont{T.~C.} \bibnamefont{{Lubensky}}},
  \bibnamefont{and} \bibinfo{author}{\bibfnamefont{J.~P.}
  \bibnamefont{{Gollub}}}, \bibinfo{journal}{Phys. Rev. E}
  \textbf{\bibinfo{volume}{65}}, \bibinfo{pages}{011307}
  (\bibinfo{year}{2002}).

\bibitem[{\citenamefont{da~Cruz et~al.}(2002)\citenamefont{da~Cruz, Chevoir,
  Bonn, and Coussot}}]{Dacruz02}
\bibinfo{author}{\bibfnamefont{F.}~\bibnamefont{da~Cruz}},
  \bibinfo{author}{\bibfnamefont{F.}~\bibnamefont{Chevoir}},
  \bibinfo{author}{\bibfnamefont{D.}~\bibnamefont{Bonn}}, \bibnamefont{and}
  \bibinfo{author}{\bibfnamefont{P.}~\bibnamefont{Coussot}},
  \bibinfo{journal}{Phys. Rev. E} \textbf{\bibinfo{volume}{66}},
  \bibinfo{pages}{051305} (\bibinfo{year}{2002}).

\bibitem[{\citenamefont{{Mueth}}(2003)}]{Mueth03}
\bibinfo{author}{\bibfnamefont{D.}~\bibnamefont{{Mueth}}},
  \bibinfo{journal}{Phys. Rev. E} \textbf{\bibinfo{volume}{67}},
  \bibinfo{pages}{011304} (\bibinfo{year}{2003}).

\bibitem[{\citenamefont{{Chambon} et~al.}(2003)\citenamefont{{Chambon},
  {Schmittbuhl}, {Corfdir}, {Vilotte}, and {Roux}}}]{Chambon03a}
\bibinfo{author}{\bibfnamefont{G.}~\bibnamefont{{Chambon}}},
  \bibinfo{author}{\bibfnamefont{J.}~\bibnamefont{{Schmittbuhl}}},
  \bibinfo{author}{\bibfnamefont{A.}~\bibnamefont{{Corfdir}}},
  \bibinfo{author}{\bibfnamefont{J.-P.} \bibnamefont{{Vilotte}}},
  \bibnamefont{and} \bibinfo{author}{\bibfnamefont{S.}~\bibnamefont{{Roux}}},
  \bibinfo{journal}{Phys. Rev. E} \textbf{\bibinfo{volume}{68}},
  \bibinfo{pages}{011304} (\bibinfo{year}{2003}).

\bibitem[{\citenamefont{Tardos et~al.}(2003)\citenamefont{Tardos, McNamara, and
  Talu}}]{Tardos03}
\bibinfo{author}{\bibfnamefont{G.}~\bibnamefont{Tardos}},
  \bibinfo{author}{\bibfnamefont{S.}~\bibnamefont{McNamara}}, \bibnamefont{and}
  \bibinfo{author}{\bibfnamefont{I.}~\bibnamefont{Talu}},
  \bibinfo{journal}{Powder Tech.} \textbf{\bibinfo{volume}{131}},
  \bibinfo{pages}{23} (\bibinfo{year}{2003}).

\bibitem[{\citenamefont{Utter and Behringer}(2004)}]{Utter04a}
\bibinfo{author}{\bibfnamefont{B.}~\bibnamefont{Utter}} \bibnamefont{and}
  \bibinfo{author}{\bibfnamefont{R.~P.} \bibnamefont{Behringer}},
  \bibinfo{journal}{Euro. Phys. J. E} \textbf{\bibinfo{volume}{14}},
  \bibinfo{pages}{373} (\bibinfo{year}{2004}).

\bibitem[{\citenamefont{da~Cruz}(2004)}]{Dacruz04a}
\bibinfo{author}{\bibfnamefont{F.}~\bibnamefont{da~Cruz}}, Ph.D. thesis,
  \bibinfo{school}{Ecole Nationale des Ponts et Chaussées}
  (\bibinfo{year}{2004}), \bibinfo{note}{in French
  (http://pastel.paristech.org/946)}.

\bibitem[{\citenamefont{Daniel et~al.}(2007)\citenamefont{Daniel, Poloski, and
  Saez}}]{Daniel07}
\bibinfo{author}{\bibfnamefont{R.}~\bibnamefont{Daniel}},
  \bibinfo{author}{\bibfnamefont{A.}~\bibnamefont{Poloski}}, \bibnamefont{and}
  \bibinfo{author}{\bibfnamefont{A.}~\bibnamefont{Saez}},
  \bibinfo{journal}{Powder Tech.} \textbf{\bibinfo{volume}{179}},
  \bibinfo{pages}{62} (\bibinfo{year}{2007}).

\bibitem[{\citenamefont{Wang et~al.}(2008)\citenamefont{Wang, Song, Briscoe,
  and Makse}}]{Wang08}
\bibinfo{author}{\bibfnamefont{P.}~\bibnamefont{Wang}},
  \bibinfo{author}{\bibfnamefont{C.}~\bibnamefont{Song}},
  \bibinfo{author}{\bibfnamefont{C.}~\bibnamefont{Briscoe}}, \bibnamefont{and}
  \bibinfo{author}{\bibfnamefont{H.}~\bibnamefont{Makse}},
  \bibinfo{journal}{Phys. Rev. E} \textbf{\bibinfo{volume}{77}},
  \bibinfo{pages}{061309} (\bibinfo{year}{2008}).

\bibitem[{\citenamefont{Nedderman}(1992)}]{Nedderman92}
\bibinfo{author}{\bibfnamefont{R.}~\bibnamefont{Nedderman}},
  \emph{\bibinfo{title}{Statics and kinematics of granular materials}}
  (\bibinfo{publisher}{Cambridge University Press},
  \bibinfo{address}{Cambridge}, \bibinfo{year}{1992}).

\bibitem[{\citenamefont{{Savage}}(1989)}]{Savage89b}
\bibinfo{author}{\bibfnamefont{S.~B.} \bibnamefont{{Savage}}}, in
  \emph{\bibinfo{booktitle}{Theoretical and applied mechanics}}, edited by
  \bibinfo{editor}{\bibfnamefont{M.}~\bibnamefont{Piau}} \bibnamefont{and}
  \bibinfo{editor}{\bibfnamefont{D.}~\bibnamefont{Caillerie}}
  (\bibinfo{publisher}{North Holland}, \bibinfo{address}{Amsterdam},
  \bibinfo{year}{1989}), pp. \bibinfo{pages}{241--266}.

\bibitem[{\citenamefont{{Chambon} et~al.}(2006)\citenamefont{{Chambon},
  {Schmittbuhl}, and {Corfdir}}}]{Chambon06b}
\bibinfo{author}{\bibfnamefont{G.}~\bibnamefont{{Chambon}}},
  \bibinfo{author}{\bibfnamefont{J.}~\bibnamefont{{Schmittbuhl}}},
  \bibnamefont{and}
  \bibinfo{author}{\bibfnamefont{A.}~\bibnamefont{{Corfdir}}},
  \bibinfo{journal}{J. Geophys. Res.} \textbf{\bibinfo{volume}{111}},
  \bibinfo{pages}{B09309} (\bibinfo{year}{2006}).

\bibitem[{\citenamefont{Karion and Hunt}(1999)}]{Karion99}
\bibinfo{author}{\bibfnamefont{A.}~\bibnamefont{Karion}} \bibnamefont{and}
  \bibinfo{author}{\bibfnamefont{M.}~\bibnamefont{Hunt}},
  \bibinfo{journal}{Journal of Heat Transfer} \textbf{\bibinfo{volume}{121}},
  \bibinfo{pages}{984} (\bibinfo{year}{1999}).

\bibitem[{\citenamefont{Schollmann}(1999)}]{Schollmann99}
\bibinfo{author}{\bibfnamefont{S.}~\bibnamefont{Schollmann}},
  \bibinfo{journal}{Phys. Rev. E} \textbf{\bibinfo{volume}{59}},
  \bibinfo{pages}{889} (\bibinfo{year}{1999}).

\bibitem[{\citenamefont{{Lätzel} et~al.}(2000)\citenamefont{{Lätzel}, {Luding},
  and {Hermann}}}]{Latzel00}
\bibinfo{author}{\bibfnamefont{M.}~\bibnamefont{{Lätzel}}},
  \bibinfo{author}{\bibfnamefont{S.}~\bibnamefont{{Luding}}}, \bibnamefont{and}
  \bibinfo{author}{\bibfnamefont{H.~J.} \bibnamefont{{Hermann}}},
  \bibinfo{journal}{Granular Matter} \textbf{\bibinfo{volume}{2}},
  \bibinfo{pages}{123} (\bibinfo{year}{2000}).

\bibitem[{\citenamefont{Zervos et~al.}(2000)\citenamefont{Zervos, Vardoulakis,
  Jean, and Lerat}}]{Zervos00}
\bibinfo{author}{\bibfnamefont{A.}~\bibnamefont{Zervos}},
  \bibinfo{author}{\bibfnamefont{I.}~\bibnamefont{Vardoulakis}},
  \bibinfo{author}{\bibfnamefont{M.}~\bibnamefont{Jean}}, \bibnamefont{and}
  \bibinfo{author}{\bibfnamefont{P.}~\bibnamefont{Lerat}},
  \bibinfo{journal}{Mech. Cohes. Frict. Mat.} \textbf{\bibinfo{volume}{5}},
  \bibinfo{pages}{305} (\bibinfo{year}{2000}).

\bibitem[{\citenamefont{Koval}(2008)}]{Koval08a}
\bibinfo{author}{\bibfnamefont{G.}~\bibnamefont{Koval}}, Ph.D. thesis,
  \bibinfo{school}{Ecole Nationale des Ponts et Chauss\'ees}
  (\bibinfo{year}{2008}), \bibinfo{note}{in French,
  http://tel.archives-ouvertes.fr/tel-00311984/fr/}.

\bibitem[{\citenamefont{Corfdir et~al.}(2004)\citenamefont{Corfdir, Lerat, and
  Vardoulakis}}]{Corfdir04}
\bibinfo{author}{\bibfnamefont{A.}~\bibnamefont{Corfdir}},
  \bibinfo{author}{\bibfnamefont{P.}~\bibnamefont{Lerat}}, \bibnamefont{and}
  \bibinfo{author}{\bibfnamefont{I.}~\bibnamefont{Vardoulakis}},
  \bibinfo{journal}{Geotechnical Testing Journal}
  \textbf{\bibinfo{volume}{27}}, \bibinfo{pages}{447} (\bibinfo{year}{2004}).

\bibitem[{\citenamefont{{Lätzel}}(2003)}]{Latzel03}
\bibinfo{author}{\bibfnamefont{M.}~\bibnamefont{{Lätzel}}}, Ph.D. thesis,
  \bibinfo{school}{University of Stuttgart} (\bibinfo{year}{2003}).

\bibitem[{\citenamefont{{Cundall} and {Strack}}(1979)}]{Cundall79}
\bibinfo{author}{\bibfnamefont{P.~A.} \bibnamefont{{Cundall}}}
  \bibnamefont{and} \bibinfo{author}{\bibfnamefont{O.~D.~L.}
  \bibnamefont{{Strack}}}, \bibinfo{journal}{G\'eotech.}
  \textbf{\bibinfo{volume}{29}}, \bibinfo{pages}{47} (\bibinfo{year}{1979}).

\bibitem[{\citenamefont{{Silbert} et~al.}(2001)\citenamefont{{Silbert},
  {Ertas}, {Grest}, {Halsey}, {Levine}, and {Plimpton}}}]{Silbert01}
\bibinfo{author}{\bibfnamefont{L.~E.} \bibnamefont{{Silbert}}},
  \bibinfo{author}{\bibfnamefont{D.}~\bibnamefont{{Ertas}}},
  \bibinfo{author}{\bibfnamefont{G.~S.} \bibnamefont{{Grest}}},
  \bibinfo{author}{\bibfnamefont{T.}~\bibnamefont{{Halsey}}},
  \bibinfo{author}{\bibfnamefont{D.}~\bibnamefont{{Levine}}}, \bibnamefont{and}
  \bibinfo{author}{\bibfnamefont{S.~J.} \bibnamefont{{Plimpton}}},
  \bibinfo{journal}{Phys. Rev. E} \textbf{\bibinfo{volume}{64}},
  \bibinfo{pages}{385} (\bibinfo{year}{2001}).

\bibitem[{\citenamefont{Roux and Chevoir}(2005)}]{Roux05}
\bibinfo{author}{\bibfnamefont{J.~N.} \bibnamefont{Roux}} \bibnamefont{and}
  \bibinfo{author}{\bibfnamefont{F.}~\bibnamefont{Chevoir}},
  \bibinfo{journal}{Bulletin des Laboratoires des Ponts et Chauss\'ees}
  \textbf{\bibinfo{volume}{254}}, \bibinfo{pages}{109} (\bibinfo{year}{2005}).

\bibitem[{\citenamefont{Rognon et~al.}(2008)\citenamefont{Rognon, Roux, Naaïm,
  and Chevoir}}]{Rognon08a}
\bibinfo{author}{\bibfnamefont{P.}~\bibnamefont{Rognon}},
  \bibinfo{author}{\bibfnamefont{J.~N.} \bibnamefont{Roux}},
  \bibinfo{author}{\bibfnamefont{M.}~\bibnamefont{Naaïm}}, \bibnamefont{and}
  \bibinfo{author}{\bibfnamefont{F.}~\bibnamefont{Chevoir}},
  \bibinfo{journal}{J. Fluid Mech.} \textbf{\bibinfo{volume}{596}},
  \bibinfo{pages}{21} (\bibinfo{year}{2008}).

\bibitem[{\citenamefont{{Allen} and {Tildesley}}(1987)}]{Allen87}
\bibinfo{author}{\bibfnamefont{M.~P.} \bibnamefont{{Allen}}} \bibnamefont{and}
  \bibinfo{author}{\bibfnamefont{D.~J.} \bibnamefont{{Tildesley}}},
  \emph{\bibinfo{title}{Computer simulation of liquids}}
  (\bibinfo{publisher}{Oxford University Press}, \bibinfo{address}{Oxford},
  \bibinfo{year}{1987}).

\bibitem[{\citenamefont{Cui et~al.}(2007)\citenamefont{Cui, O’Sullivan, and
  O’Neill}}]{Cui07}
\bibinfo{author}{\bibfnamefont{L.}~\bibnamefont{Cui}},
  \bibinfo{author}{\bibfnamefont{C.}~\bibnamefont{O’Sullivan}},
  \bibnamefont{and} \bibinfo{author}{\bibfnamefont{S.}~\bibnamefont{O’Neill}},
  \bibinfo{journal}{G\'eotech.} \textbf{\bibinfo{volume}{57}},
  \bibinfo{pages}{831–} (\bibinfo{year}{2007}).

\bibitem[{\citenamefont{{Campbell}}(2002)}]{Campbell02}
\bibinfo{author}{\bibfnamefont{C.~S.} \bibnamefont{{Campbell}}},
  \bibinfo{journal}{J. Fluid Mech.} \textbf{\bibinfo{volume}{465}},
  \bibinfo{pages}{261} (\bibinfo{year}{2002}).

\bibitem[{\citenamefont{{Combe}}(2002)}]{Combe02a}
\bibinfo{author}{\bibfnamefont{G.}~\bibnamefont{{Combe}}},
  \emph{\bibinfo{title}{Microscopic origins of strain in granular materials}},
  vol. \bibinfo{volume}{SI8} (\bibinfo{publisher}{Collection Etudes et
  Recherches des Laboratoires des Ponts et Chaussées},
  \bibinfo{address}{Paris}, \bibinfo{year}{2002}), \bibinfo{note}{in French
  (http://pastel.paristech.org/51/)}.

\bibitem[{\citenamefont{Radjaï and Roux}(2004)}]{Radjai04}
\bibinfo{author}{\bibfnamefont{F.}~\bibnamefont{Radjaï}} \bibnamefont{and}
  \bibinfo{author}{\bibfnamefont{S.}~\bibnamefont{Roux}}, in
  \emph{\bibinfo{booktitle}{The physics of granular media}}, edited by
  \bibinfo{editor}{\bibfnamefont{H.}~\bibnamefont{Hinrichsen}}
  \bibnamefont{and} \bibinfo{editor}{\bibfnamefont{D.}~\bibnamefont{Wolf}}
  (\bibinfo{publisher}{Wiley-Vch}, \bibinfo{address}{Weinheim},
  \bibinfo{year}{2004}), pp. \bibinfo{pages}{165--187}.

\bibitem[{\citenamefont{Radjaï}(2008)}]{Radjai08}
\bibinfo{author}{\bibfnamefont{F.}~\bibnamefont{Radjaï}}
  (\bibinfo{year}{2008}), \bibinfo{note}{arXiv:0801.4722}.

\bibitem[{\citenamefont{Coleman et~al.}(1966)\citenamefont{Coleman, Markovitz,
  and Noll}}]{Coleman66}
\bibinfo{author}{\bibfnamefont{B.}~\bibnamefont{Coleman}},
  \bibinfo{author}{\bibfnamefont{H.}~\bibnamefont{Markovitz}},
  \bibnamefont{and} \bibinfo{author}{\bibfnamefont{W.}~\bibnamefont{Noll}},
  \emph{\bibinfo{title}{Viscosimetric flows of non {N}ewtonian fluids}}
  (\bibinfo{publisher}{Springer-Verlag}, \bibinfo{address}{Berlin},
  \bibinfo{year}{1966}).

\bibitem[{\citenamefont{{Savage} and {Sayed}}(1984)}]{Savage84}
\bibinfo{author}{\bibfnamefont{S.~B.} \bibnamefont{{Savage}}} \bibnamefont{and}
  \bibinfo{author}{\bibfnamefont{M.}~\bibnamefont{{Sayed}}},
  \bibinfo{journal}{J. Fluid Mech.} \textbf{\bibinfo{volume}{142}},
  \bibinfo{pages}{391} (\bibinfo{year}{1984}).

\bibitem[{\citenamefont{Losert et~al.}(2000)\citenamefont{Losert, Bocquet,
  Lubensky, and Gollub}}]{Losert00a}
\bibinfo{author}{\bibfnamefont{W.}~\bibnamefont{Losert}},
  \bibinfo{author}{\bibfnamefont{L.}~\bibnamefont{Bocquet}},
  \bibinfo{author}{\bibfnamefont{T.~C.} \bibnamefont{Lubensky}},
  \bibnamefont{and} \bibinfo{author}{\bibfnamefont{J.}~\bibnamefont{Gollub}},
  \bibinfo{journal}{Phys. Rev. Lett.} \textbf{\bibinfo{volume}{85}},
  \bibinfo{pages}{1428} (\bibinfo{year}{2000}).

\bibitem[{\citenamefont{Denniston and Li}(1999)}]{Denniston99}
\bibinfo{author}{\bibfnamefont{C.}~\bibnamefont{Denniston}} \bibnamefont{and}
  \bibinfo{author}{\bibfnamefont{H.}~\bibnamefont{Li}}, \bibinfo{journal}{Phys.
  Rev. E} \textbf{\bibinfo{volume}{59}}, \bibinfo{pages}{3289}
  (\bibinfo{year}{1999}).

\bibitem[{\citenamefont{{Ertas} et~al.}(2001)\citenamefont{{Ertas}, {Grest},
  {Halsey}, {Levine}, and {Silbert}}}]{Ertas01}
\bibinfo{author}{\bibfnamefont{D.}~\bibnamefont{{Ertas}}},
  \bibinfo{author}{\bibfnamefont{G.~S.} \bibnamefont{{Grest}}},
  \bibinfo{author}{\bibfnamefont{T.~C.} \bibnamefont{{Halsey}}},
  \bibinfo{author}{\bibfnamefont{D.}~\bibnamefont{{Levine}}}, \bibnamefont{and}
  \bibinfo{author}{\bibfnamefont{L.}~\bibnamefont{{Silbert}}},
  \bibinfo{journal}{Europhys. Lett.} \textbf{\bibinfo{volume}{56}},
  \bibinfo{pages}{214} (\bibinfo{year}{2001}).

\bibitem[{\citenamefont{Chevoir et~al.}(2001)\citenamefont{Chevoir, Prochnow,
  Jenkins, and Mills}}]{Chevoir01a}
\bibinfo{author}{\bibfnamefont{F.}~\bibnamefont{Chevoir}},
  \bibinfo{author}{\bibfnamefont{M.}~\bibnamefont{Prochnow}},
  \bibinfo{author}{\bibfnamefont{J.}~\bibnamefont{Jenkins}}, \bibnamefont{and}
  \bibinfo{author}{\bibfnamefont{P.}~\bibnamefont{Mills}}, in
  \emph{\bibinfo{booktitle}{Powders and grains 2001}}, edited by
  \bibinfo{editor}{\bibfnamefont{Y.}~\bibnamefont{Kishino}}
  (\bibinfo{publisher}{Balkema}, \bibinfo{address}{Rotterdam},
  \bibinfo{year}{2001}), pp. \bibinfo{pages}{373--376}.

\bibitem[{\citenamefont{Mills et~al.}(2008)\citenamefont{Mills, Rognon, and
  Chevoir}}]{Mills08}
\bibinfo{author}{\bibfnamefont{P.}~\bibnamefont{Mills}},
  \bibinfo{author}{\bibfnamefont{P.}~\bibnamefont{Rognon}}, \bibnamefont{and}
  \bibinfo{author}{\bibfnamefont{F.}~\bibnamefont{Chevoir}},
  \bibinfo{journal}{Europhys. Lett.} \textbf{\bibinfo{volume}{81}},
  \bibinfo{pages}{64005} (\bibinfo{year}{2008}).

\bibitem[{\citenamefont{Tejchman and Wu}(1993)}]{Tejchman93b}
\bibinfo{author}{\bibfnamefont{J.}~\bibnamefont{Tejchman}} \bibnamefont{and}
  \bibinfo{author}{\bibfnamefont{W.}~\bibnamefont{Wu}}, \bibinfo{journal}{Acta
  Mechanica} \textbf{\bibinfo{volume}{99}}, \bibinfo{pages}{61}
  (\bibinfo{year}{1993}).

\bibitem[{\citenamefont{{Mohan} et~al.}(2002)\citenamefont{{Mohan}, {Rao}, and
  {Nott}}}]{Mohan02}
\bibinfo{author}{\bibfnamefont{L.~S.} \bibnamefont{{Mohan}}},
  \bibinfo{author}{\bibfnamefont{K.~K.} \bibnamefont{{Rao}}}, \bibnamefont{and}
  \bibinfo{author}{\bibfnamefont{P.~R.} \bibnamefont{{Nott}}},
  \bibinfo{journal}{J. Fluid Mech.} \textbf{\bibinfo{volume}{457}},
  \bibinfo{pages}{377} (\bibinfo{year}{2002}).

\bibitem[{\citenamefont{Artoni et~al.}(2008)\citenamefont{Artoni, Canu, and
  Santomaso}}]{Artoni08}
\bibinfo{author}{\bibfnamefont{R.}~\bibnamefont{Artoni}},
  \bibinfo{author}{\bibfnamefont{P.}~\bibnamefont{Canu}}, \bibnamefont{and}
  \bibinfo{author}{\bibfnamefont{A.}~\bibnamefont{Santomaso}}
  (\bibinfo{year}{2008}), \bibinfo{note}{arXiv:0803.3890}.

\bibitem[{\citenamefont{Rajchenbach}(2000)}]{Rajchenbach00}
\bibinfo{author}{\bibfnamefont{J.}~\bibnamefont{Rajchenbach}},
  \bibinfo{journal}{Adv. in Physics} \textbf{\bibinfo{volume}{49}},
  \bibinfo{pages}{229} (\bibinfo{year}{2000}).

\bibitem[{\citenamefont{Khakhar et~al.}(2001)\citenamefont{Khakhar, Orpe,
  Andresen, and Ottino}}]{Khakhar01}
\bibinfo{author}{\bibfnamefont{D.}~\bibnamefont{Khakhar}},
  \bibinfo{author}{\bibfnamefont{A.}~\bibnamefont{Orpe}},
  \bibinfo{author}{\bibfnamefont{P.}~\bibnamefont{Andresen}}, \bibnamefont{and}
  \bibinfo{author}{\bibfnamefont{J.}~\bibnamefont{Ottino}},
  \bibinfo{journal}{J. Fluid Mech.} \textbf{\bibinfo{volume}{441}},
  \bibinfo{pages}{255} (\bibinfo{year}{2001}).

\bibitem[{\citenamefont{Komatsu et~al.}(2001)\citenamefont{Komatsu, Inagaki,
  Nakagawa, and Nasuno}}]{Komatsu01}
\bibinfo{author}{\bibfnamefont{T.~S.} \bibnamefont{Komatsu}},
  \bibinfo{author}{\bibfnamefont{S.}~\bibnamefont{Inagaki}},
  \bibinfo{author}{\bibfnamefont{M.}~\bibnamefont{Nakagawa}}, \bibnamefont{and}
  \bibinfo{author}{\bibfnamefont{S.}~\bibnamefont{Nasuno}},
  \bibinfo{journal}{Phys. Rev. Lett.} \textbf{\bibinfo{volume}{86}},
  \bibinfo{pages}{1757} (\bibinfo{year}{2001}).

\bibitem[{\citenamefont{Crassous et~al.}(2008)\citenamefont{Crassous, Metayer,
  Richard, and Laroche}}]{Crassous08}
\bibinfo{author}{\bibfnamefont{J.}~\bibnamefont{Crassous}},
  \bibinfo{author}{\bibfnamefont{J.-F.} \bibnamefont{Metayer}},
  \bibinfo{author}{\bibfnamefont{P.}~\bibnamefont{Richard}}, \bibnamefont{and}
  \bibinfo{author}{\bibfnamefont{C.}~\bibnamefont{Laroche}},
  \bibinfo{journal}{J. Stat. Mech.}  (\bibinfo{year}{2008}),
  \bibinfo{note}{doi: 10.1088/1742-5468/2008/03/P03009}.

\bibitem[{\citenamefont{Debrégeas and Josserand}(2000)}]{Debregeas00}
\bibinfo{author}{\bibfnamefont{G.}~\bibnamefont{Debrégeas}} \bibnamefont{and}
  \bibinfo{author}{\bibfnamefont{C.}~\bibnamefont{Josserand}},
  \bibinfo{journal}{Europhys. Lett.} \textbf{\bibinfo{volume}{52}},
  \bibinfo{pages}{137} (\bibinfo{year}{2000}).

\bibitem[{\citenamefont{Lagree and Lhuillier}(2006)}]{Lagree06}
\bibinfo{author}{\bibfnamefont{P.-Y.} \bibnamefont{Lagree}} \bibnamefont{and}
  \bibinfo{author}{\bibfnamefont{D.}~\bibnamefont{Lhuillier}},
  \bibinfo{journal}{Eur. J. Mech. B} \textbf{\bibinfo{volume}{25}},
  \bibinfo{pages}{960} (\bibinfo{year}{2006}).

\bibitem[{\citenamefont{Falk et~al.}(2008)\citenamefont{Falk, Toiya, and
  Losert}}]{Falk08}
\bibinfo{author}{\bibfnamefont{M.~L.} \bibnamefont{Falk}},
  \bibinfo{author}{\bibfnamefont{M.}~\bibnamefont{Toiya}}, \bibnamefont{and}
  \bibinfo{author}{\bibfnamefont{W.}~\bibnamefont{Losert}}
  (\bibinfo{year}{2008}), \bibinfo{note}{arXiv:0802.1752}.

\bibitem[{\citenamefont{Debr\'egeas et~al.}(2001)\citenamefont{Debr\'egeas,
  Tabuteau, and di~Meglio}}]{Debregeas01}
\bibinfo{author}{\bibfnamefont{G.}~\bibnamefont{Debr\'egeas}},
  \bibinfo{author}{\bibfnamefont{H.}~\bibnamefont{Tabuteau}}, \bibnamefont{and}
  \bibinfo{author}{\bibfnamefont{J.-M.} \bibnamefont{di~Meglio}},
  \bibinfo{journal}{Phys. Rev. Lett.} \textbf{\bibinfo{volume}{87}},
  \bibinfo{pages}{178305} (\bibinfo{year}{2001}).

\bibitem[{\citenamefont{{Huang} et~al.}(2005)\citenamefont{{Huang}, {Ovarlez},
  {Bertrand}, {Rodts}, {Coussot}, and {Bonn}}}]{Huang05}
\bibinfo{author}{\bibfnamefont{N.}~\bibnamefont{{Huang}}},
  \bibinfo{author}{\bibfnamefont{G.}~\bibnamefont{{Ovarlez}}},
  \bibinfo{author}{\bibfnamefont{F.}~\bibnamefont{{Bertrand}}},
  \bibinfo{author}{\bibfnamefont{S.}~\bibnamefont{{Rodts}}},
  \bibinfo{author}{\bibfnamefont{P.}~\bibnamefont{{Coussot}}},
  \bibnamefont{and} \bibinfo{author}{\bibfnamefont{D.}~\bibnamefont{{Bonn}}},
  \bibinfo{journal}{Phys. Rev. Lett.} \textbf{\bibinfo{volume}{94}},
  \bibinfo{pages}{028301} (\bibinfo{year}{2005}).

\bibitem[{\citenamefont{Moreau}(1997)}]{Moreau97}
\bibinfo{author}{\bibfnamefont{J.-J.} \bibnamefont{Moreau}}, in
  \emph{\bibinfo{booktitle}{Friction, Arching, Contact Dynamics}}, edited by
  \bibinfo{editor}{\bibfnamefont{D.~E.} \bibnamefont{{Wolf}}} \bibnamefont{and}
  \bibinfo{editor}{\bibfnamefont{P.}~\bibnamefont{{Grassberger}}}
  (\bibinfo{publisher}{World Scientific}, \bibinfo{address}{Londres},
  \bibinfo{year}{1997}), pp. \bibinfo{pages}{233--247}.

\bibitem[{\citenamefont{Staron et~al.}(2002)\citenamefont{Staron, Vilotte, and
  Radjaï}}]{Staron02b}
\bibinfo{author}{\bibfnamefont{L.}~\bibnamefont{Staron}},
  \bibinfo{author}{\bibfnamefont{J.-P.} \bibnamefont{Vilotte}},
  \bibnamefont{and} \bibinfo{author}{\bibfnamefont{F.}~\bibnamefont{Radjaï}},
  \bibinfo{journal}{Phys. Rev. Lett.} \textbf{\bibinfo{volume}{89}},
  \bibinfo{pages}{204302} (\bibinfo{year}{2002}).

\bibitem[{\citenamefont{Radjaï and Roux}(2002)}]{Radjai02}
\bibinfo{author}{\bibfnamefont{F.}~\bibnamefont{Radjaï}} \bibnamefont{and}
  \bibinfo{author}{\bibfnamefont{S.}~\bibnamefont{Roux}},
  \bibinfo{journal}{Phys. Rev. Lett.} \textbf{\bibinfo{volume}{89}},
  \bibinfo{pages}{064302} (\bibinfo{year}{2002}).

\end{thebibliography}
\end{document}